\documentclass{aa}  

\usepackage{graphicx}
\usepackage{txfonts}
\usepackage{natbib,twoopt,morefloats}
\usepackage[breaklinks=true]{hyperref} 
\bibpunct{(}{)}{;}{a}{}{,}             
\makeatletter
  \newcommandtwoopt{\citeads}[3][][]{\href{http://adsabs.harvard.edu/abs/#3}%
    {\def\hyper@linkstart##1##2{}%
     \let\hyper@linkend\@empty\citealp[#1][#2]{#3}}}
  \newcommandtwoopt{\citepads}[3][][]{\href{http://adsabs.harvard.edu/abs/#3}%
    {\def\hyper@linkstart##1##2{}%
     \let\hyper@linkend\@empty\citep[#1][#2]{#3}}}
  \newcommandtwoopt{\citetads}[3][][]{\href{http://adsabs.harvard.edu/abs/#3}%
    {\def\hyper@linkstart##1##2{}%
     \let\hyper@linkend\@empty\citet[#1][#2]{#3}}}
  \newcommandtwoopt{\citeyearads}[3][][]%
    {\href{http://adsabs.harvard.edu/abs/#3}
    {\def\hyper@linkstart##1##2{}%
     \let\hyper@linkend\@empty\citeyear[#1][#2]{#3}}}
\makeatother

\usepackage{color}

\begin{document}

   \title{Astrometry of the main satellites of Uranus: 18 years of observations\thanks{Based on observations made at the
Pico dos Dias Observatory/LNA.}
\thanks{Tables with the positions of Uranus and its satellites, as well as with the X and Y CCD coordinates of the observed satellites
and reference stars, are available at the CDS via anonymous ftp to 
cdsarc.u-strasbg.fr (130.79.128.5) or via http://cdsarc.u-strasbg.fr/viz-bin/qcat?J/A+A/???/A???}}
\titlerunning{Astrometry of Uranus and of its main satellites}


   \author{J.~I.~B.~Camargo\inst{1,2}
          \and
           F.~P.~Magalh\~aes\inst{1,3}
          \and
           R.~Vieira-Martins\inst{1,2,4,5}
          \and
           M.~Assafin\inst{5,6}
          \and
          F.~Braga-Ribas\inst{1,7}
          \and
          A.~Dias-Oliveira\inst{1}
          \and
          G.~Benedetti-Rossi\inst{1}
          \and
          A.~R.~Gomes-J\'unior\inst{5}
         \and 
          A.~H.~Andrei\inst{1,5,8,9}
          \and
          D.~N.~da Silva Neto\inst{10}
           }

   \institute{Observat\'orio Nacional/MCTI, R. General Jos\'e Cristino 77, CEP 20921-400 
              Rio de Janeiro, RJ, Brazil\\
              \email{camargo@on.br}
        \and
             Laborat\'orio Interinstitucional de e-Astronomia - LIneA, Rua Gal. Jos\'e Cristino 77, 
                  Rio de Janeiro, RJ 20921-400, Brazil
        \and
             Instituto Nacional de Pesquisas Espaciais, Av. dos Astronautas 1758, CEP 12227-010
             S\~ao Jos\'e dos Campos, SP, Brazil
        \and
             Affiliated researcher at the Observatoire de Paris/IMCCE, 77 av. Denfert-Rochereau,
             75014 Paris, France
        \and
             Observat\'orio do Valongo/UFRJ, Ladeira do Pedro Ant\^onio 43, CEP 20080-090 
             Rio de Janeiro, RJ, Brazil
        \and
             Associated researcher at the Observatoire de Paris/IMCCE, 77 av. Denfert-Rochereau,
             75014 Paris, France
        \and
             Federal University of Technology - Paran\'a (UTFPR / DAFIS), Rua Sete de Setembro, 3165
             CEP 80230-901, Curitiba, PR, Brazil
        \and
             Associated researcher at the Observatoire de Paris/SYRTE, 77 av. Denfert Rochereau, 
             75014 Paris, France
        \and
             Associated researcher at the INAF/Osservatorio Astronomico di Torino, Strada Osservatorio 
             20, 10025 Pino Torinesi (To), Italy
        \and 
             Centro Universit\'ario Estadual da Zona Oeste, Av. Manual Caldeira de Alvarenga 1203, CEP 
             23.070-200 Rio de Janeiro RJ, Brazil
             }

   \date{Received; accepted}

 
  \abstract
{We contribute to developing dynamical models of the motions of Uranus' main satellites.}
   {We determine accurate positions of the main satellites of Uranus: Miranda, Ariel, Umbriel, Titania, and
Oberon. Positions of Uranus, as derived from those of these satellites, are also determined. The observational
period spans from 1992 to 2011. All runs were made at the Pico dos Dias Observatory, Brazil.}
   {We used the software called Platform for Reduction of Astronomical Images Automatically (PRAIA) to perform a
digital coronography to minimise the influence of the scattered light of Uranus on the astrometric measurements 
and to determine accurate positions of the main satellites. The
 positions of Uranus were then indirectly determined by computing
the mean differences between the observed and ephemeris positions of these satellites.
A series of numerical filters was applied to filter out spurious data. These filters are mostly based on
(a) the comparison between the positions of Oberon with those of the other satellites and on (b) the 
offsets as given by the differences between the observed and ephemeris positions of all satellites.
}
   { We have, for the overall offsets of the five satellites,
$-$29 mas ($\pm$63 mas) in right ascension and $-$27 mas ($\pm$46 mas) in declination. For the overall 
difference between the offsets of Oberon and those of the other satellites, we have $+$3 mas ($\pm$30 mas) in right ascension and $-$2 mas ($\pm$28 mas) in declination.
Ephemeris positions for the satellites were determined from DE432$+$ura111. Comparisons using other 
modern ephemerides for the solar system -- INPOP13c -- and for the motion of the satellites -- 
NOE-7-2013 -- were also made. They confirm that the largest contribution to the offsets we find comes from the motion of the 
barycenter of the Uranus system around the barycenter of the solar system, as given by the planetary ephemerides. 
For the period from 1992 to 2011, our final catalogues contain 584 observed positions of Miranda, 1,710 
of Ariel, 1,987 of Umbriel, 2,588 of Titania, 2,928 of Oberon, and 3,516 of Uranus.}
   {}

   \keywords{Astrometry -- 
                    Methods: data analysis -- 
                    Catalogs -- 
                    Planets and satellites: individual: Uranus, Miranda, Ariel, Umbriel, Titania, Oberon
               }

   \maketitle
%

\section{Introduction}

Accurate positions from gound-based CCD images of small bodies and natural satellites in the
solar system are an important tool for developing accurate dynamical models of their motions 
\citepads{2008P&SS...56.1766L} and to support \citepads[see][for instance]{2015A&A..XXX..XXXD} further 
investigations of these bodies and 
their surrounding environments by other observational techniques and 
methods, such as stellar occultations \citepads[][]{2014Natur.508...72B}. In 
addition, the knowledge of the dynamics of natural satellites is important 
to constrain models of formation and evolution of the solar system \citepads[][]{2011Icar..214..113N},
to understand their internal structure \citepads[][]{2009Natur.459..957L}, and to more general
studies of the physics of a particular planetary system \citepads[see][for the Uranus system]{2014AJ....148...76J}.

A summary of ground-based techniques used in the astrometry of solar system bodies,
along with their respective accuracies, can be found in \citetads{2008IAUS..248...66A}.
Stellar occultations can also be included among those techniques since they are also
a source of accurate positions \citepads[see][]{2006AJ....132.1575P,
2006Natur.439...52S,2009Icar..199..458W,2010A&A...515A..32A,2014A&A...570A..86B}.
In the case of Pluto, for instance, one could take advantage of the fact that its photocenter 
displacement due to the presence of Charon is not of concern for the astrometry from 
stellar occultations.

Here, we aim at contributing to dynamical models of the motions of the main satellites of Uranus: 
Miranda, Ariel, Umbriel, Titania, and Oberon. This contribution is made by determining 
accurate positions determined from CCD observations carried out at the Pico dos Dias Observatory in 
the period 1992-2011. 

Uranus was not directly observed in this work. However, positions of Uranus, 
indirectly determined from those of the satellites, are also provided since 
they
are relevant to improve its orbit. In fact, orbits in modern planetary 
ephemerides are not uniformly known. Those of Uranus, Neptune, and Pluto have
uncertainties of several thousand kilometers, two orders of magnitude greater 
than those for  Jupiter and Saturn. The situation is even more dramatic when the comparison is made with
the inner solar system, where orbits are known to subkilometric accuracy
\citepads[see][for further details]{2014IPNPR.196C...1F}. 

This work represents an extension of that by \citetads{2003AJ....125.2714V} (hereafter V03) and
follows the efforts of many others that determined optical positions within the system of Uranus from 
the ground. In addition to traditional astrometry, 
a noticeable contribution comes also from the observation of mutual events, where \citetads{2008MNRAS.384L..38H}
were the first to report such an analysis to our knowledge that involved the satellites of Uranus. Comprehensive lists
of observational sources for the system of Uranus can be found, for instance, in \citetads{2014AJ....148...76J}
and \citet{2013MNRAS.436.3668E}.

Next, in Sect.~2, we describe the site and instruments. Data reduction procedure and the results and analysis 
are presented in Sects.~3 and 4, respectively. Comparisons with other ephemerides are given in
Sect.~5. Section~6 argues the extension of the astrometry of this 
work to older epochs using results from V03, and Sect.~7 presents our position catalogues. Comments and 
conclusions are given in Sect.~8.


\section{Site and instruments}

All observations were made with three telescopes located at the Pico dos Dias Observatory (IAU code 874, 
$\lambda=-45^{\rm o} 34^{\prime} 57^{\prime\prime}$~W, $\phi=-22^{\rm o} 32^{\prime} 04^{\prime\prime}$,
and h=1810.7m),
run by the Laborat\'orio Nacional de Astrof\'{\i}sica/MCTI\footnote{http://www.lna.br/. Page in Portuguese.}.
Their characteristics are summarised in Table~\ref{table1}. All of them have equatorial mounts.
\begin{table}
\caption{Characteristics of the telescopes}             
\label{table1}      
\begin{center}          
\begin{tabular}{l c c c c }     
\hline\hline       
Telescope               &       f       &       f/      &       scale                           & System      \\
                        &       (mm)    &               & ($^{\prime\prime}$/mm)                &               \\
\hline
Perkin-Elmer            &       15752   &       f/10    &       13.09                           & RC          \\
Boller\&Chivens         &       8222    &       f/13.49 &       25.09                           & RC          \\
Zeiss                   &       7500    &       f/12.5  &       27.50                           & Cass                \\
\hline\hline       
\end{tabular}
\end{center}          
Columns are manufacturer, focal length, f-number, pixel scale, and telescope
system (Ritchey-Chr\'etien -- RC or Cassegrain -- Cass).
\end{table}

A number of detectors, as given by Table~\ref{table2}, were installed
at the telescopes presented
in Table~\ref{table1}. Typically, our images from the Perkin-Elmer telescope have a field
of view (FOV)
of 3$^{\prime}$ to 6$^{\prime}$ , and for the other telescopes, the FOV is 5$^{\prime}$ to 
11$^{\prime}$.

\begin{table}
\caption{CCD detectors}             
\label{table2}      
\begin{center}          
\begin{tabular}{l c c c}     
\hline\hline       
Type            &       Size (pixels)                   &       Size (microns)                       &       $\lambda$ (nm)  \\
\hline
(a)             &       2048  $\times$ 4608     &       13.5  $\times$ 13.5          &       500                             \\
(b)             &       2048  $\times$ 2048     &       13.5  $\times$ 13.5          &       600                             \\
(c)             &       1024  $\times$ 1024     &       24    $\times$ 24            &       650                             \\
(d)             &       2048  $\times$ 2048     &       13.5  $\times$ 13.5          &       491                             \\
(e)             &       1024  $\times$ 1024     &       13.0  $\times$ 13.0  &       560                             \\
\hline\hline       
\end{tabular}
\end{center}     
CCD types are (a) Marconi CCD42-90-0-941, (b) Marconi CCD42-40-1-368, (c) SITe SI003AB, 
(d) Andor iKon-L, and (e) Andor iXon$^{\rm EM}$. The other columns give the CCD dimensions,
pixel sizes, and the wavelength of the CCD maximum quantum efficiency.
\end{table}

Filters {\it B}, {\it V}, {\it R}, and {\it I}, compatible with the Johnsons-Cousins system, were used. 
Most of our observations, however, were made without filter.

\section{Data reduction}

All raw images were first corrected for bias and flatfield with IRAF \citepads{1993ASPC...52..173T}. 
Then, a digital coronography procedure \citepads[see][]{2008P&SS...56.1882A, 2009AJ....137.4046A} was applied to 
minimise the effect of the scattered light from Uranus on the positions of the satellites.

   \begin{figure}
   \centering{
   \includegraphics[width=4.3cm]{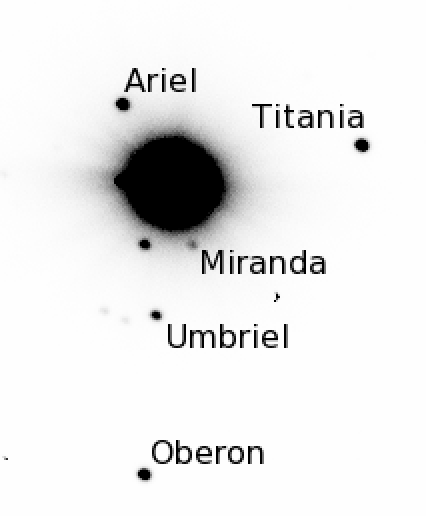}\hspace {10pt}\includegraphics[width=4.3cm]{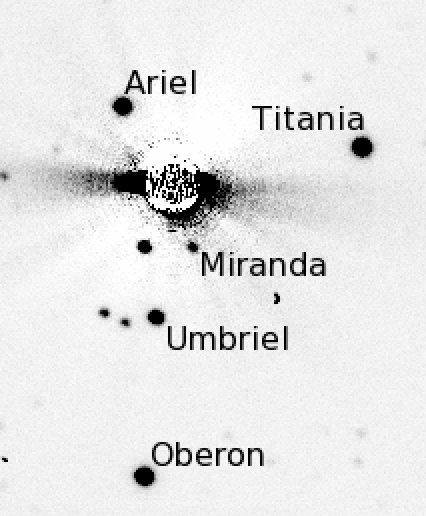}}
   \vskip 5pt
   \includegraphics[width=4.3cm]{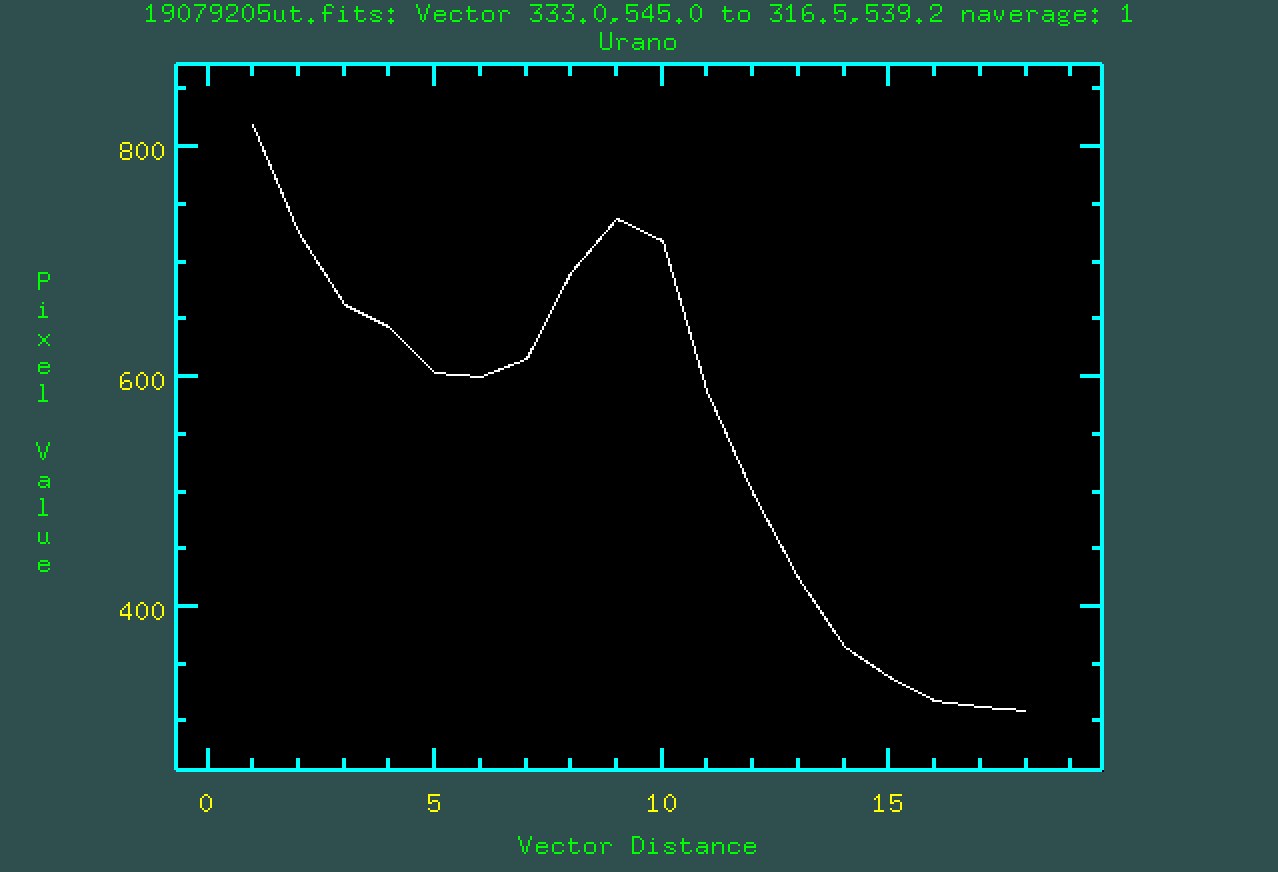}\hspace{10pt}\includegraphics[width=4.3cm]{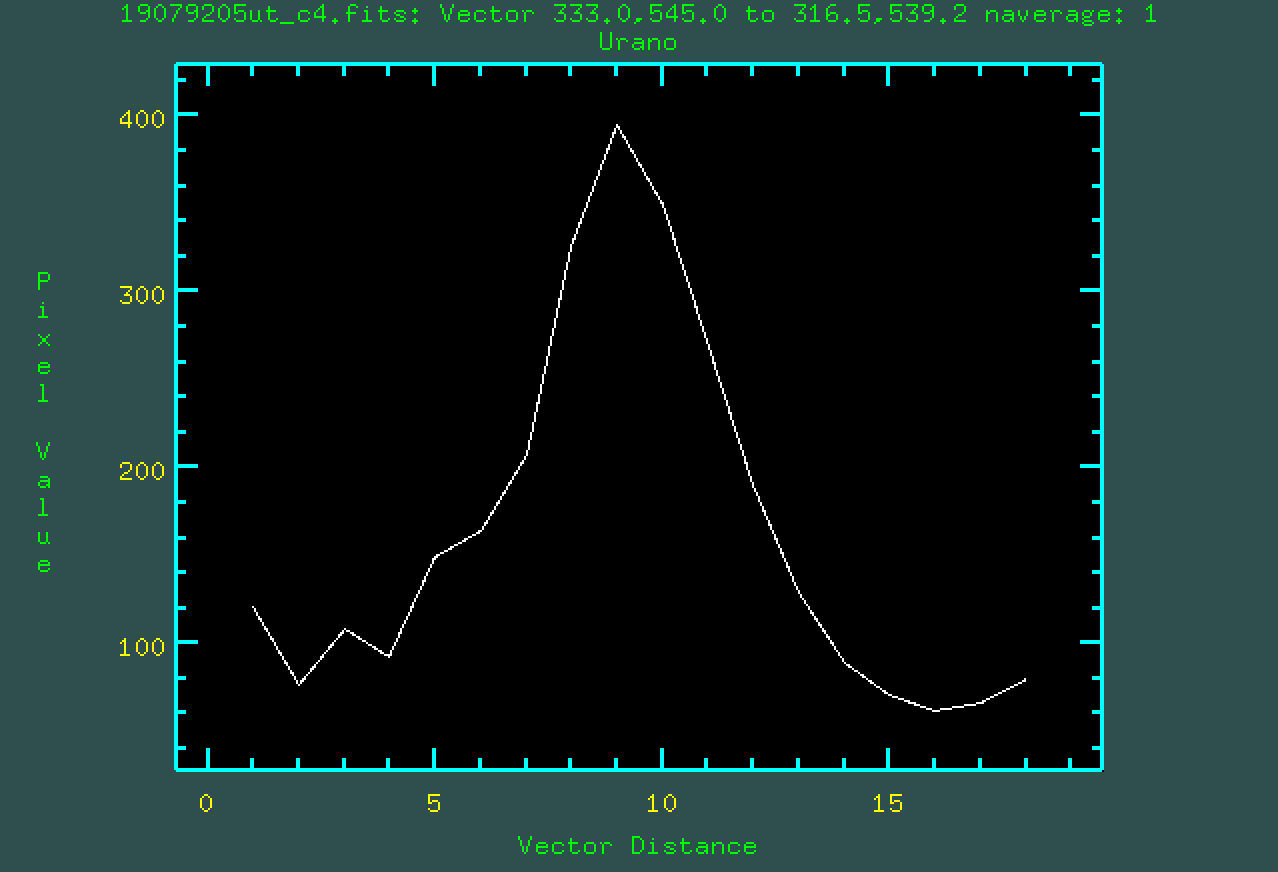}      
\caption{Image containing Uranus and its main satellites, as obtained on 18$^{\rm }$ July, 1992,
with the Perkin-Elmer telescope. Upper left (right) panel: inverted colour map image before (after) 
the application of the digital coronography procedure. Lower left (right) panel: respective 
Miranda flux profile along a segment placed on the line joining the centres of Uranus and Miranda. 
This segment is the same in the two lower panels. Image orientation is north up and east left. Exposure time
is 9s. All other objects in the image that do not belong to the system of Uranus 
are field stars.
             }
         \label{figure1}
   \end{figure}

Figure~\ref{figure1} illustrates the performance of this correction. In this figure, the angular separation between 
the centres of Uranus and Miranda is $9.5^{\prime\prime}$. Although the centroid determination of any object
contaminated by the scattered light of Uranus profits from the digital coronography, this procedure 
is evidently more relevant to Miranda in our images. 

The final step, that is, the determination of the positions of the satellites of Uranus in all images,  was 
fully implemented with the Platform for Reduction of Astronomical Images Automatically (PRAIA)
package\footnote{The digital coronography procedure is also part of this package.} 
\citepads[][]{2011gfun.conf...85A}. 

\begin{figure}
   \centering{
   \includegraphics[width=9cm]{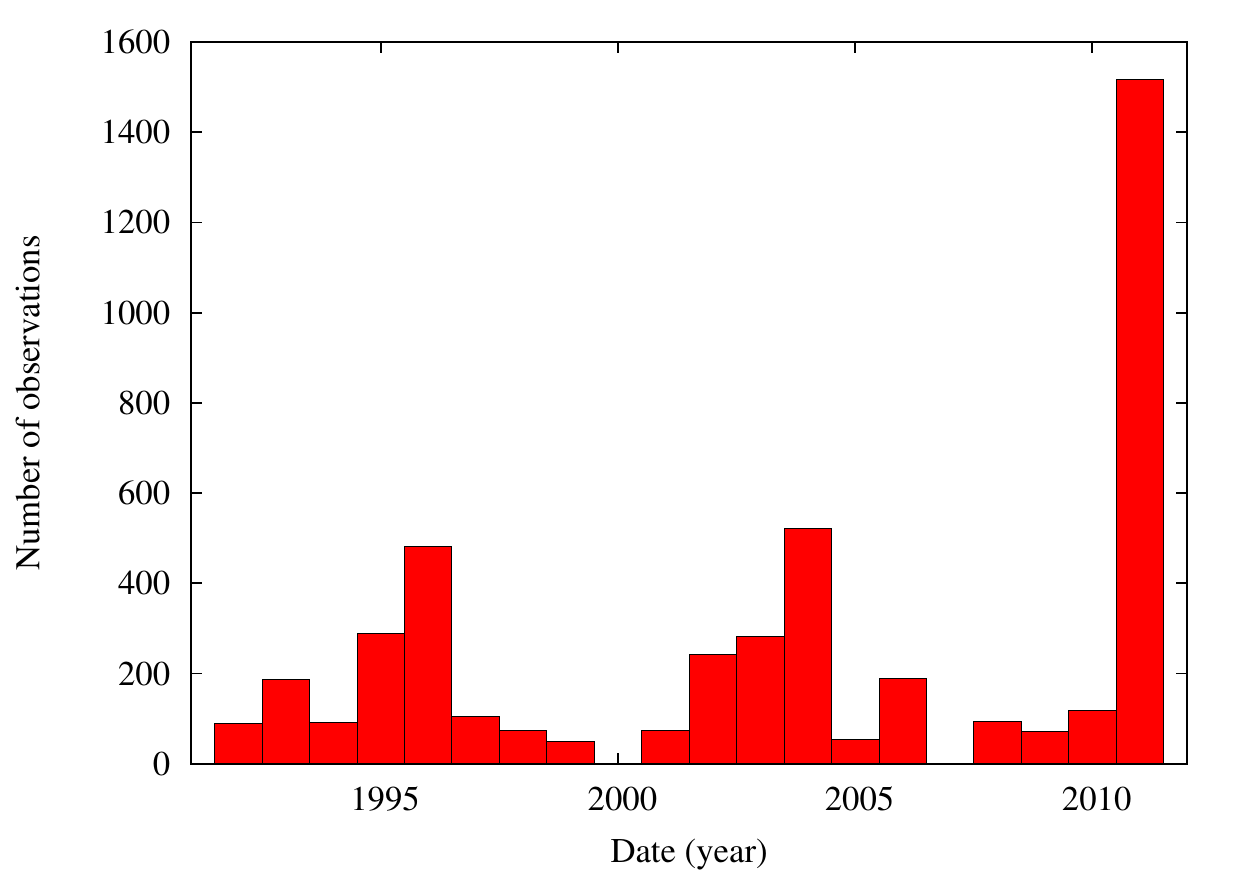}}
\caption{Distribution of the number of observations per year.
             }
         \label{figure2}
   \end{figure}

The results presented here were obtained from
the analysis of a total of 4,532 images, all telescopes and detectors shown in Tables~\ref{table1} 
and \ref{table2} included, distributed along 138 nights in the period 1992-2011 (see also Fig.~\ref{figure2}). These counts exclude 
images (23\% of the total) where positions, for whatever reason, were not obtained. We used the 
Fourth US Naval Observatory CCD Astrograph Catalog \citepads[UCAC4 -- ][]{2013AJ....145...44Z}
as the practical representative of the International 
Celestial Reference System \citepads[ICRS -- ][]{1995A&A...303..604A}.

Figure~\ref{figure2} also shows the absence of observations in 2000 and 2007. In
2000, we had no observational runs at the Pico dos Dias Observatory. In 2007, all observations (more 
than 18,000) of the Uranus system were aimed at determining dynamical parameters from light curves 
obtained during mutual events among the satellites \citepads[see][]{2009AJ....137.4046A}. As a result of their 
short exposure times (1-3 seconds) and small FOV (usually 2$^{\prime}$$\times$2$^{\prime}$ or less), 
they were not used in this work.

\section{Results and data analysis}

The observed positions of the satellites were first compared to those provided
by ephemerides from the JPL: the planetary ephemeris DE432 and 
ura111\footnote{both at http://naif.jpl.nasa.gov/pub/naif/generic\_kernels/spk/}. 

By means of the SPICE system toolkits \citepads{1996P&SS...44...65A}, DE432 was used to
link the geocenter to the barycenter of the Uranus system, whereas ura111 was used to link this
latter to the centre of Uranus and those of its main satellites. The link between the geocenter
and the topocenter was obtained with subroutines from the Standards of
Fundamental Astronomy\footnote{http://www.iausofa.org} (SOFA) and from the Naval Observatory Vector 
Astrometry Software\footnote{http://aa.usno.navy.mil/software/novas/novas\_info.php} (NOVAS).
The sum of these links (or vectors) produced the astrometric ephemeris positions of Uranus and of its 
main satellites for an observer at the Pico dos Dias Observatory. 

The differences in the sense observed minus ephemeris positions are referred to as\textup{\textup{\textup{\textup{\textup{\textup{
{\it \textup{offsets}}}}}}}} in position, following the nomenclature commonly used in astrometric works. 
These offsets were determined and 
used to filter out spurious data. Details on this filtering are given below. We note that, except for
item 1, no further consideration was made with respect to the telescope, filter, or detector used.

\begin{enumerate}
\item All satellite positions whose centres were closer than $8^{\prime\prime}$ (for observations made 
with the Perkin-Elmer telescope) or $7^{\prime\prime}$ (for observations made with the Boller\&Chivens or 
Zeiss telescopes) to the centre of Uranus were discarded.
This was done to avoid cases where the digital coronography might fail to produce reliable results.
\item All images with astrometry derived from less than five reference (UCAC4) stars were discarded.
This was done avoid a poor representation of the celestial frame.
\item For each night and each satellite, an iterative $\sigma$-clipping procedure 
filtered out all those offsets that differed from the mean offset in right ascension or declination
by more than 2-$\sigma$; this way, we eliminated poor measurements.
\item To all observations of a given satellite, an iterative $\sigma$-clipping procedure filtered out all 
those residuals in right ascension and declination, as given by the difference between the offsets of the given 
satellite and those of Oberon in the same respective images that differed from the mean residual by 
more than 3-$\sigma$. This was a coherence check, assuming that the orbits of the satellites around the 
barycenter of the Uranus system are well known. This is
discussed below. Oberon was chosen as reference because its positions were determined in 
all the 4,532 images mentioned earlier and, most relevant, it is the most distant and the slowest among 
the satellites studied here (as a consequence, Oberon has the most realiable orbit of the satellites
studied here).
\item To all observations of a given satellite, an iterative $\sigma$-clipping procedure filtered out all 
those offsets in right ascension and declination that differed from the mean offset by more than 
3-$\sigma$. We did this to use the information from the whole data set to filter out poor representations 
of the celestial frame that survived item 2 and that can only be identified when a large portion of
the sky is considered. 
\end{enumerate}

These filters were applied one after the other in the sequence they have just been presented. The most severe
of them, that is, the one that eliminated the most, was that described in item 3. The least severe procedure
was that described in item 5.

\begin{figure}
   \centering{
   \includegraphics[width=9cm]{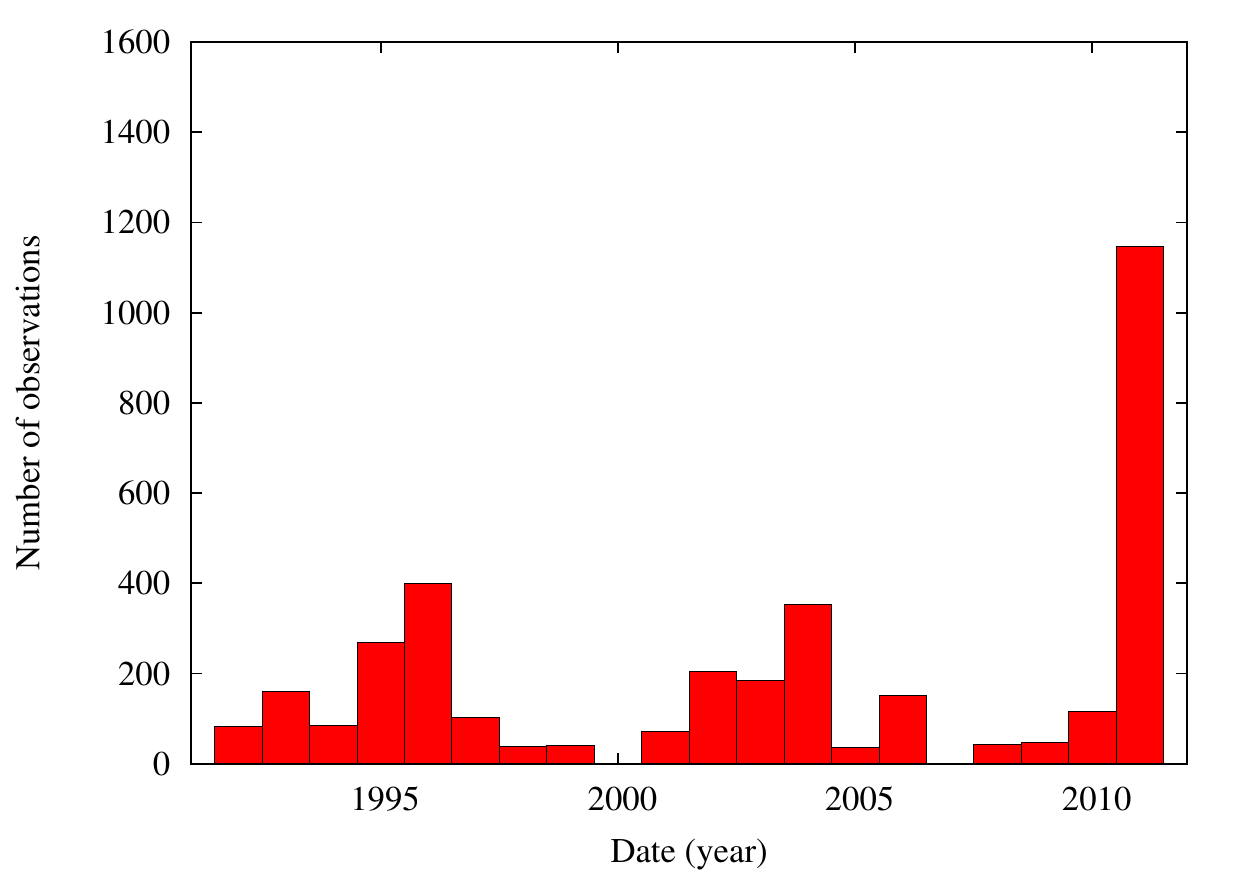}}
\caption{Distribution of the number of observations per year after applying
filtres 1 to 5. 
             }
         \label{figure3}
   \end{figure}

\begin{figure}
   \centering{
   \includegraphics[width=4.5cm]{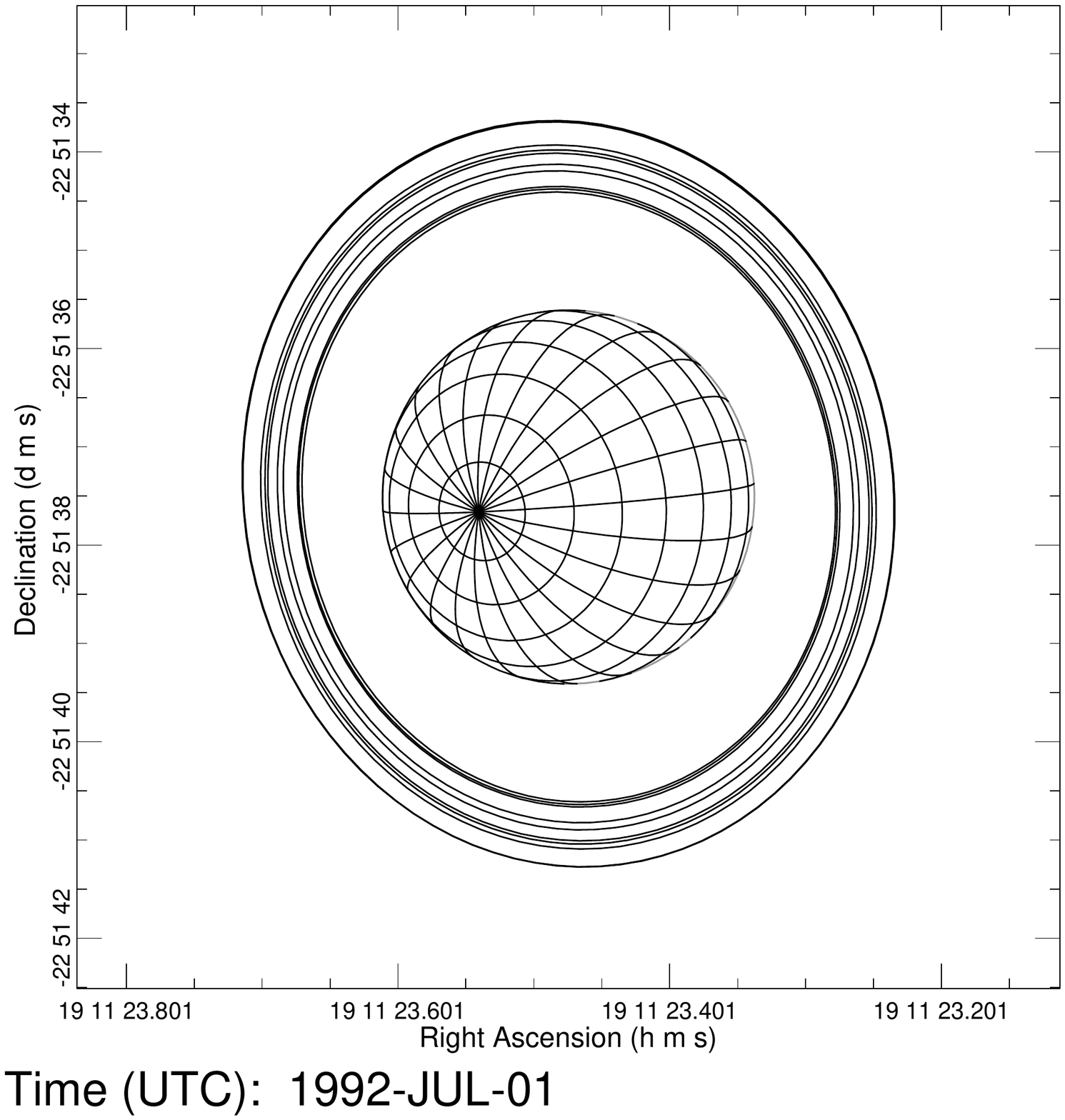}\includegraphics[width=4.5cm]{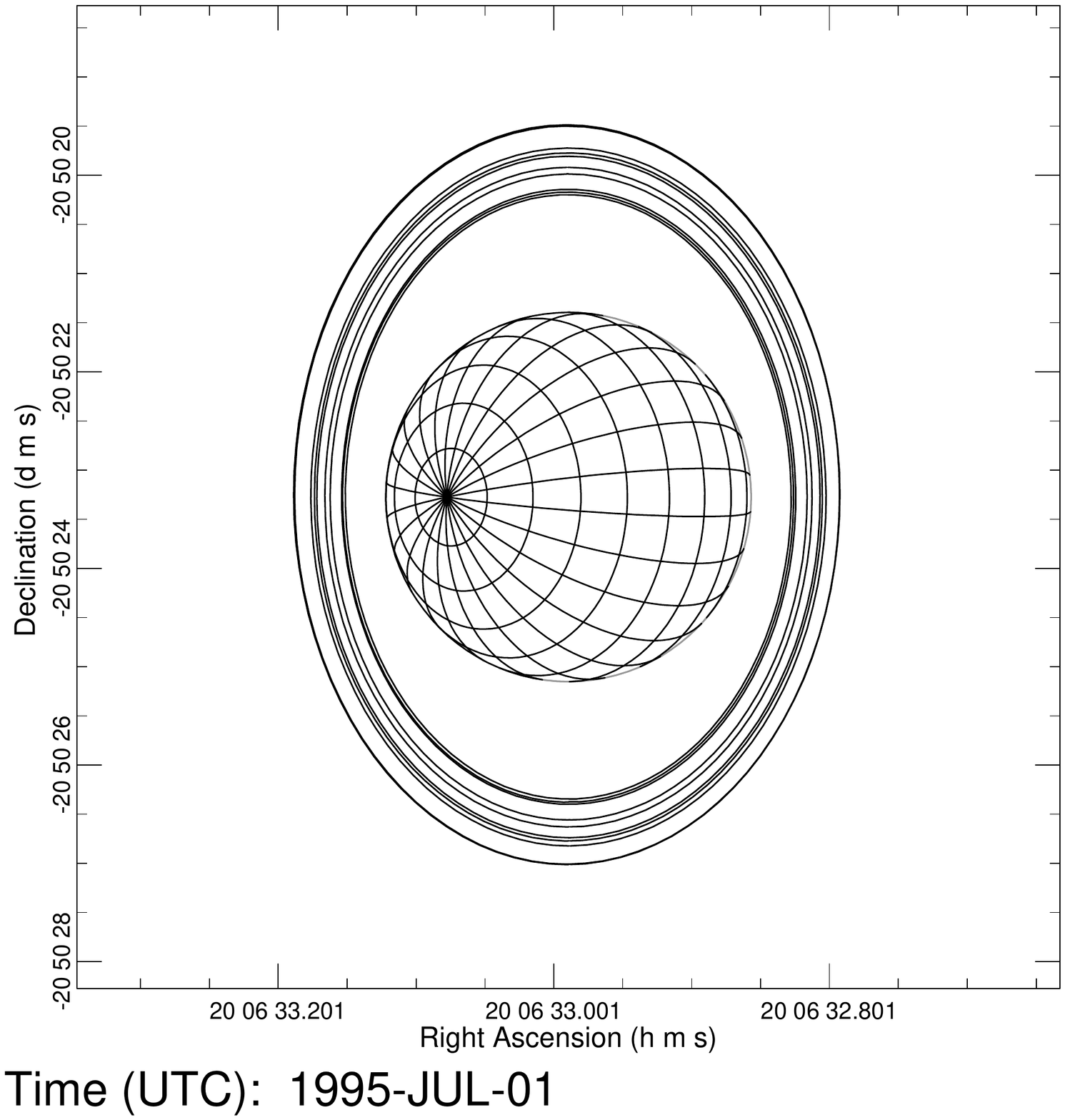}
   \includegraphics[width=4.5cm]{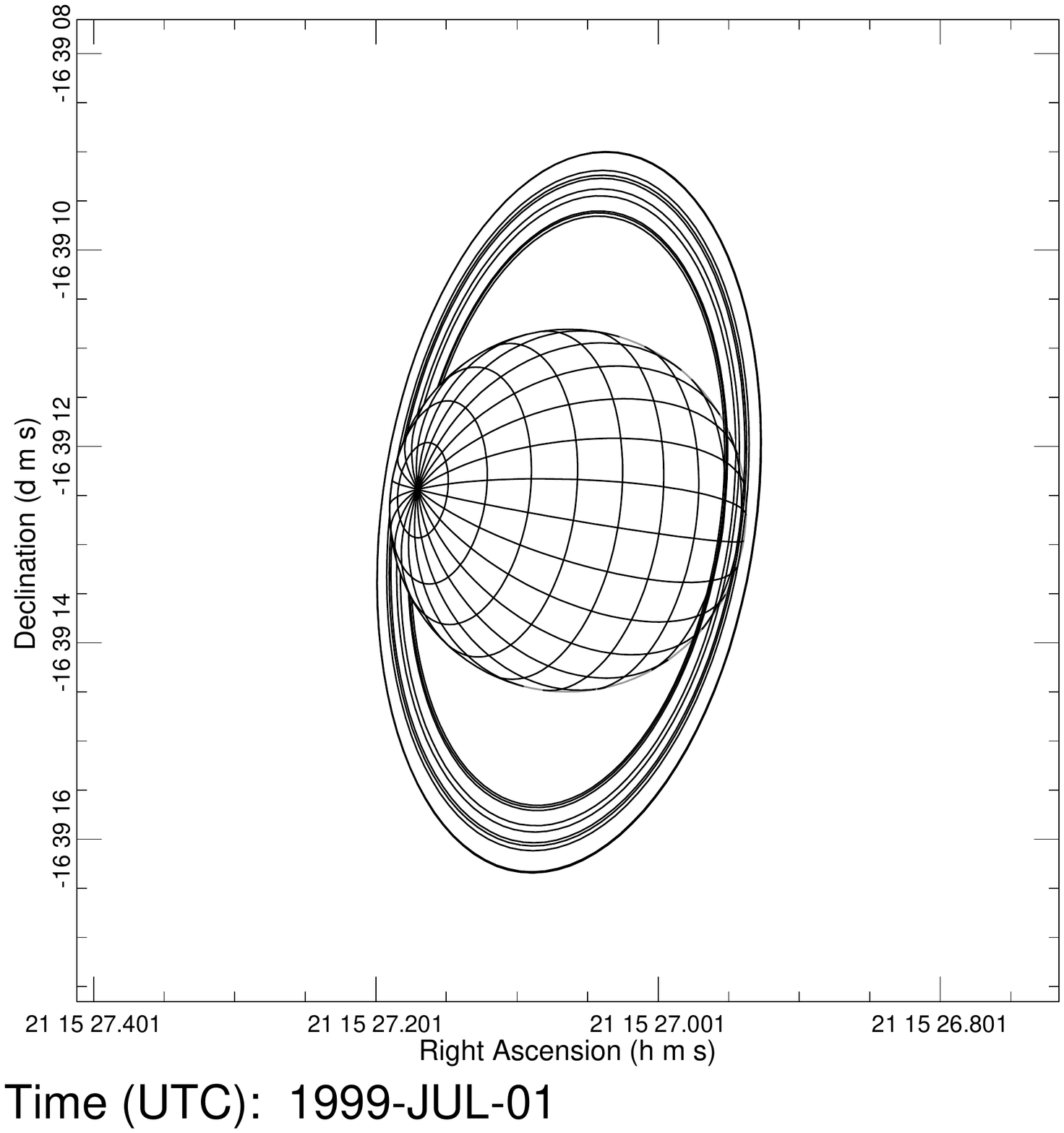}\includegraphics[width=4.5cm]{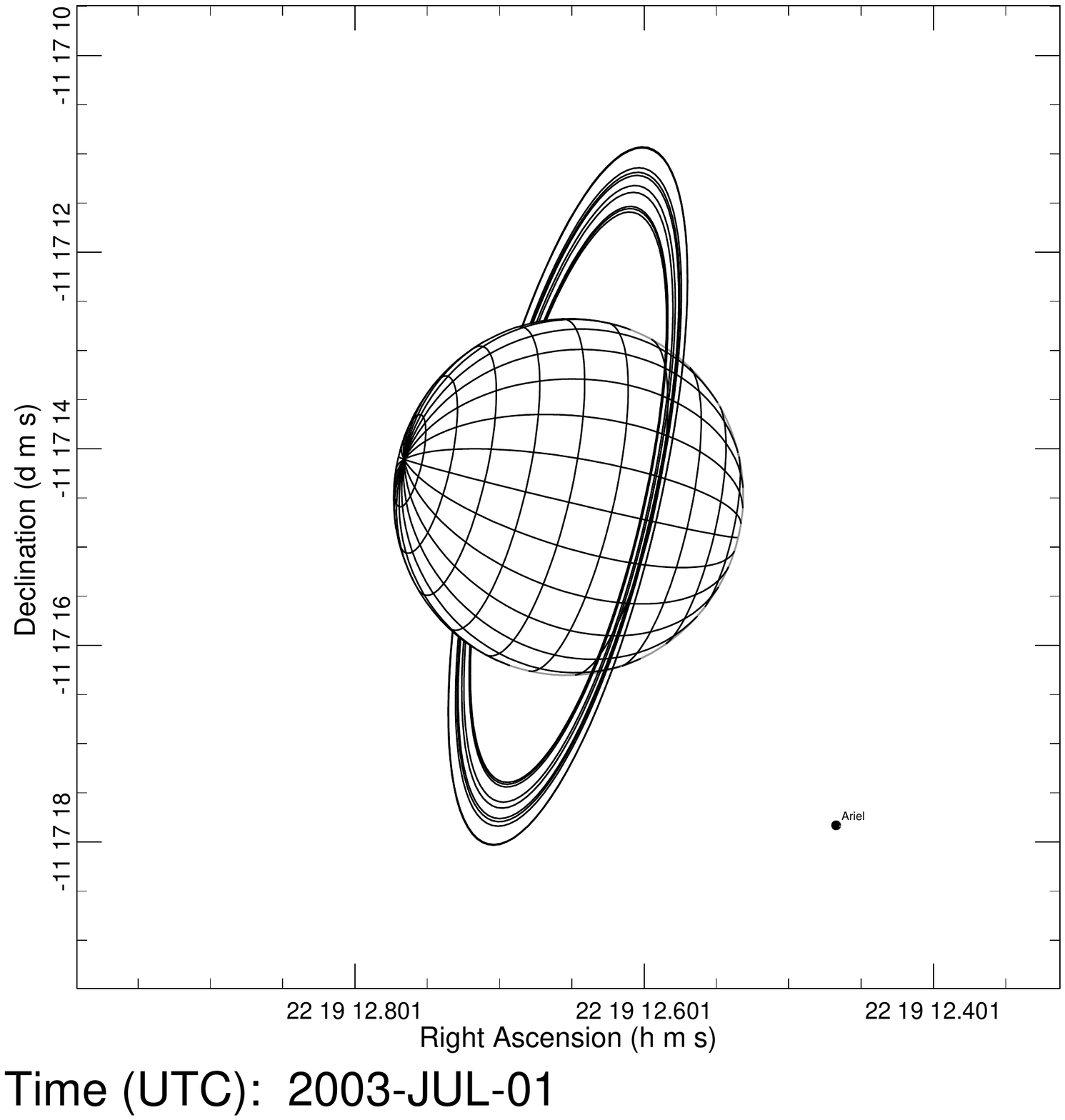}
   \includegraphics[width=4.5cm]{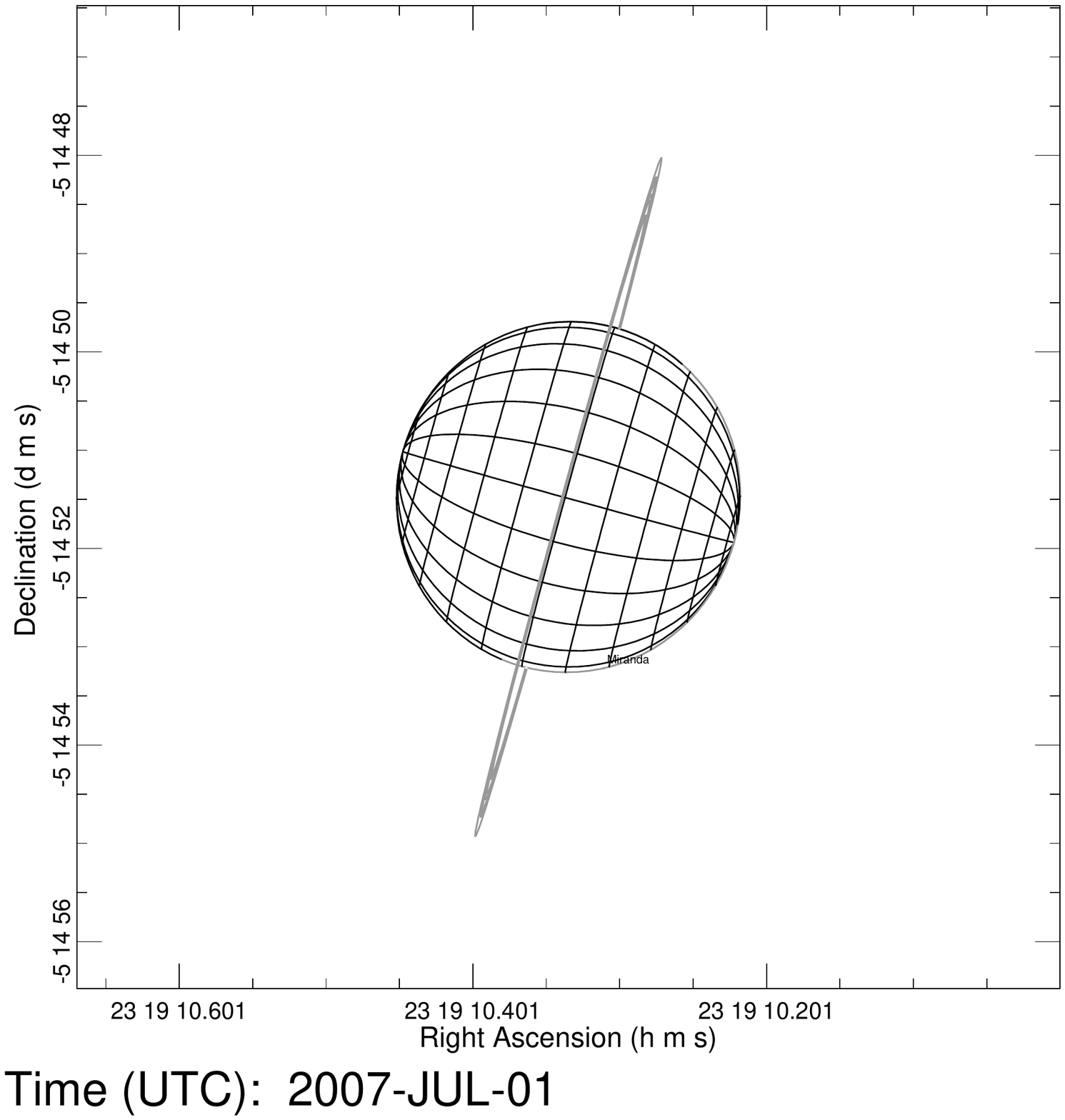}\includegraphics[width=4.5cm]{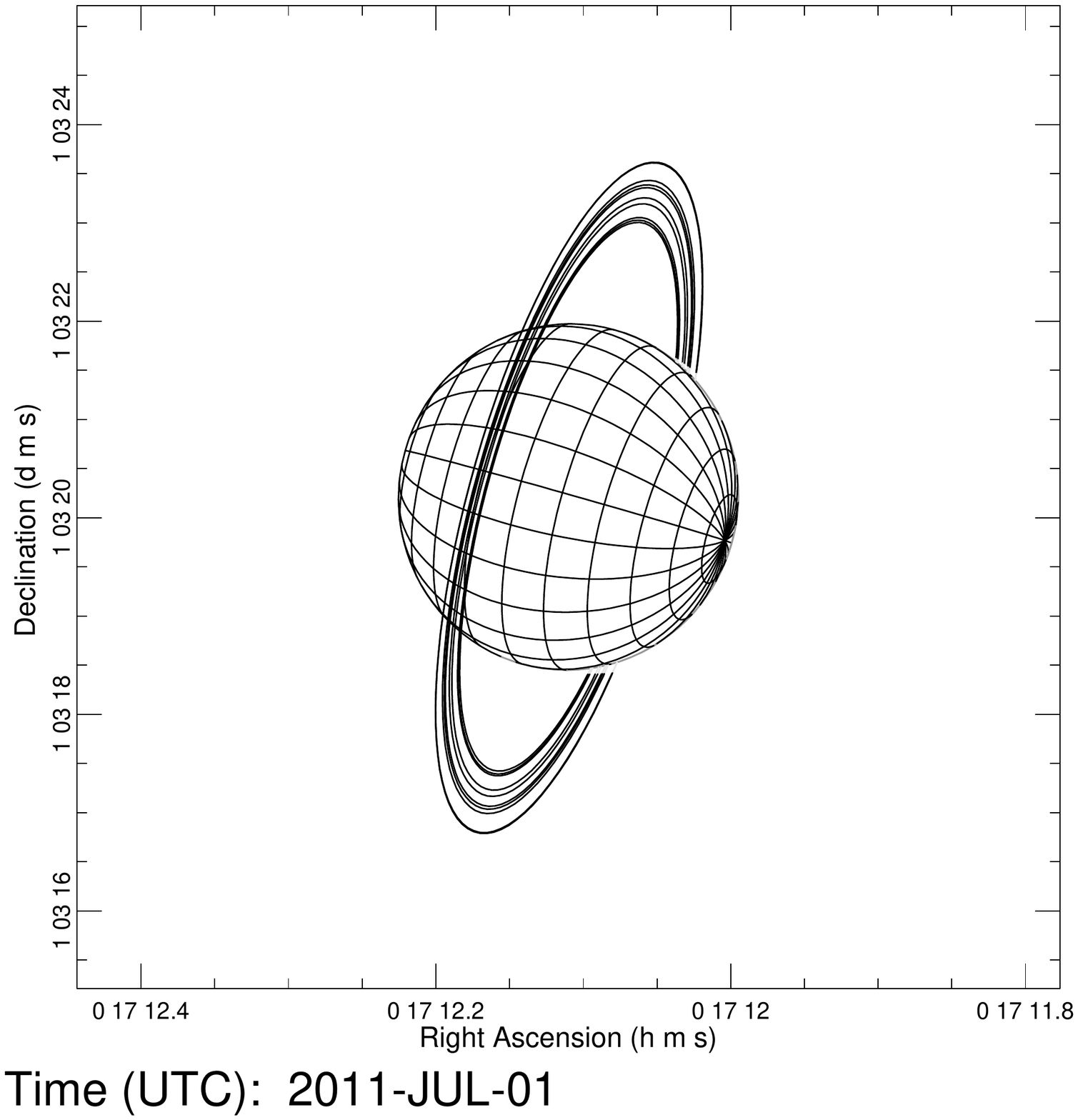}
             }
\caption{Evolution of the pole position of Uranus, with the view$^{\ref{ft1}}$ of the satellite's orbit planes
gradually changing from close to face-on (1992) to edge-on (2007).
             }
         \label{figure4}
   \end{figure}

Figure~\ref{figure3} shows the number of observations per year after applying all filters.
From comparing Figs.~\ref{figure2} and \ref{figure3}, it is clear that positions derived from observations 
made after 2000 were more affected by the application of the filters than those made earlier. The reason
for this can be easily explained by Fig.~\ref{figure4}. After 2000, the planes of the orbits of the satellites were approaching an
edge-on view so that observations became more complicated, and
in addition, more satellite positions were filtered 
out because of their proximity to Uranus.

\begin{figure}
   \centering{
   \includegraphics[width=6.5cm]{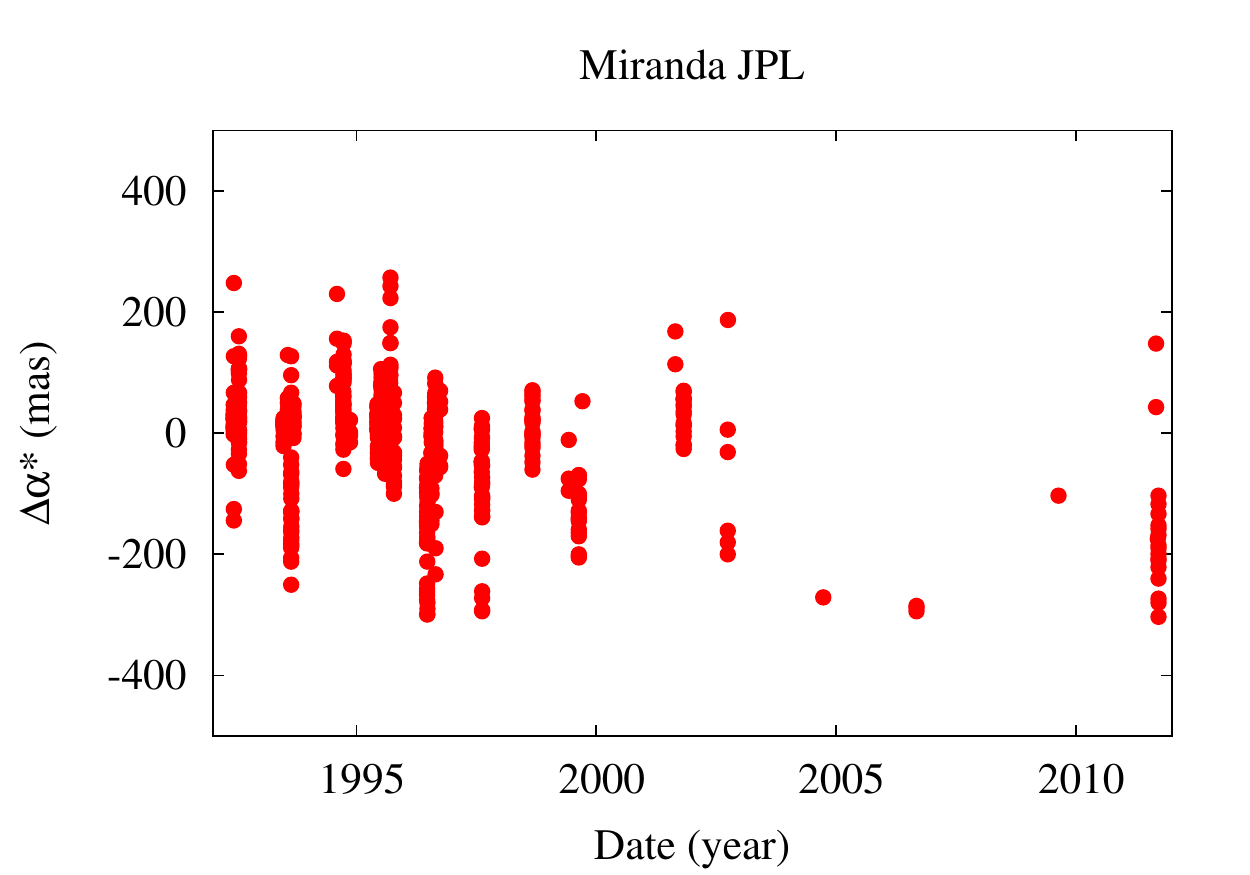}
   \includegraphics[width=6.5cm]{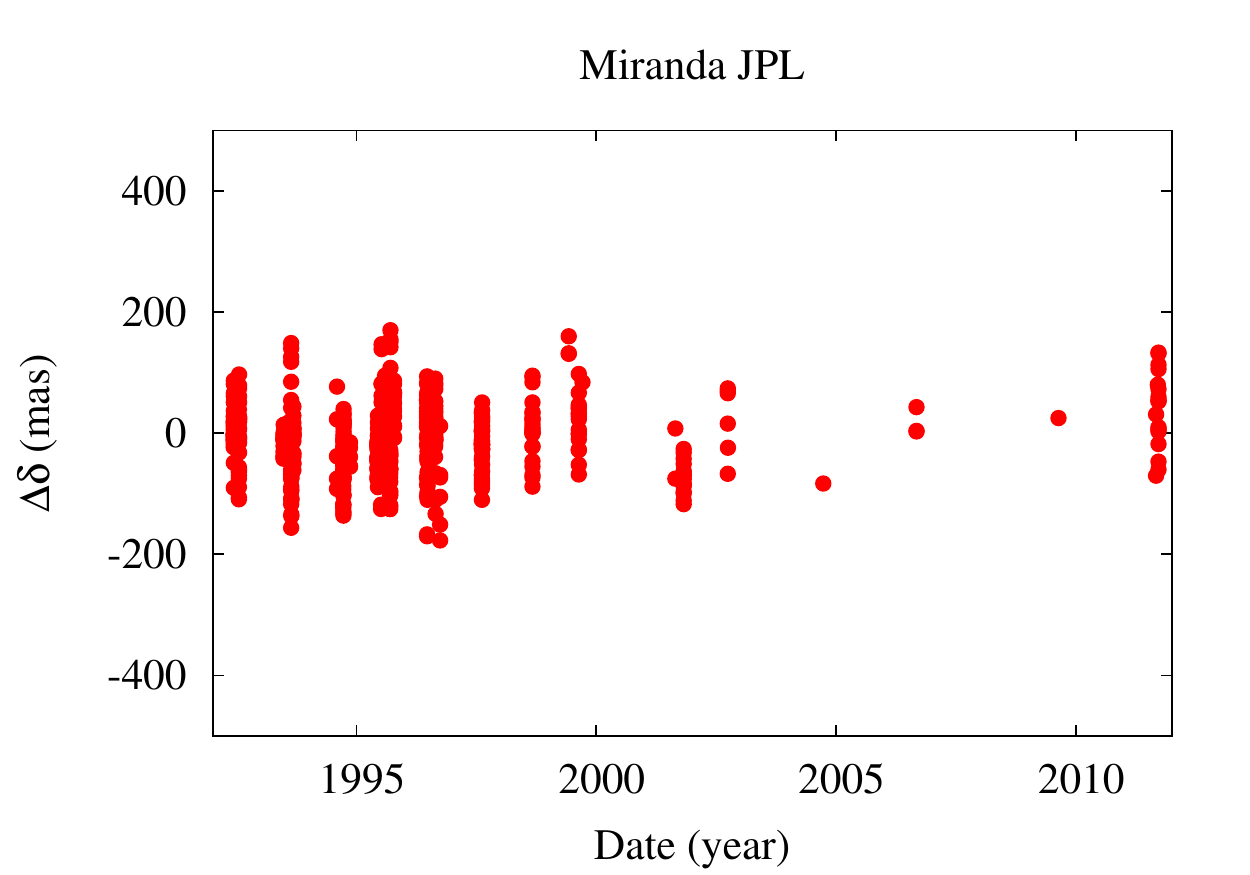}}
\caption{Distribution of the offsets for Miranda as a function of time.
             }
         \label{figure5}
   \end{figure}

\begin{figure}
   \centering{
   \includegraphics[width=6.5cm]{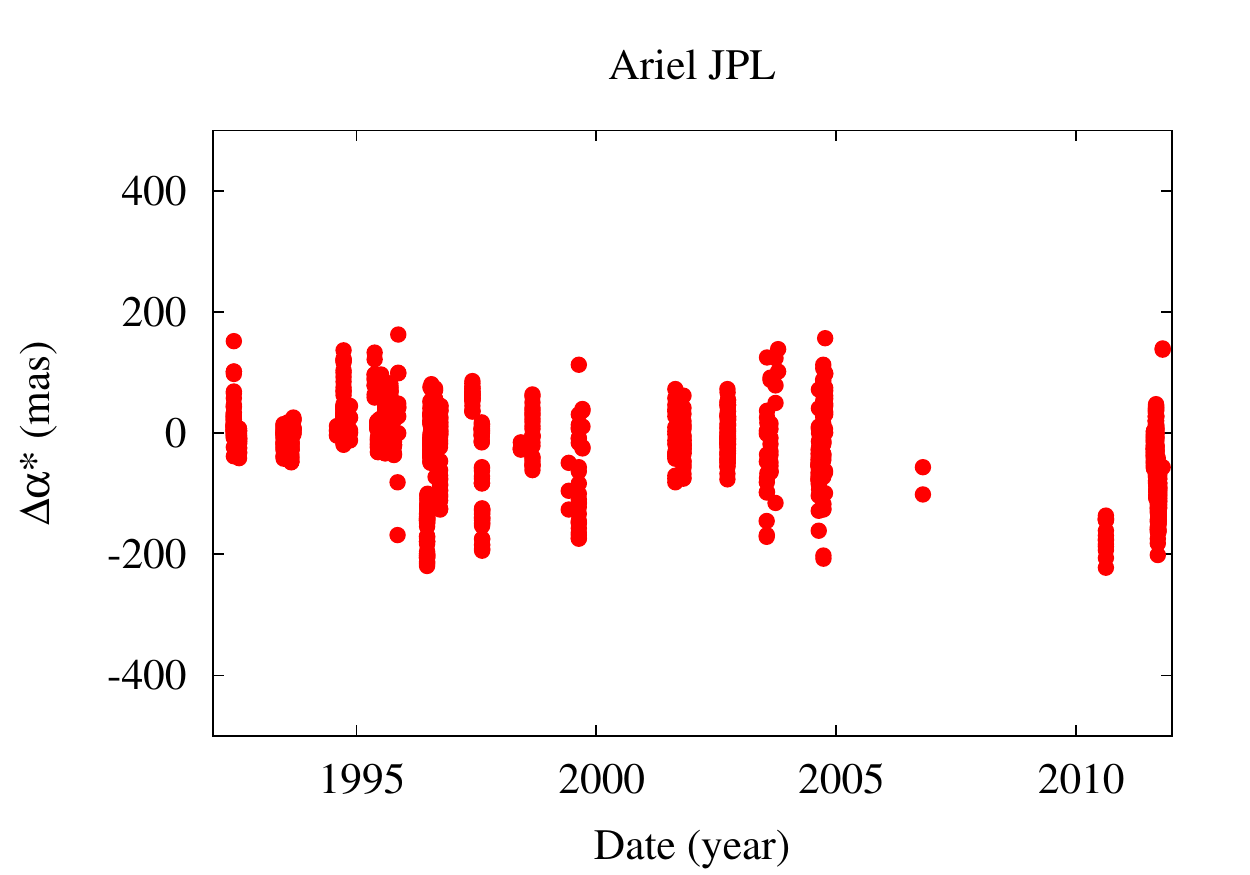}
   \includegraphics[width=6.5cm]{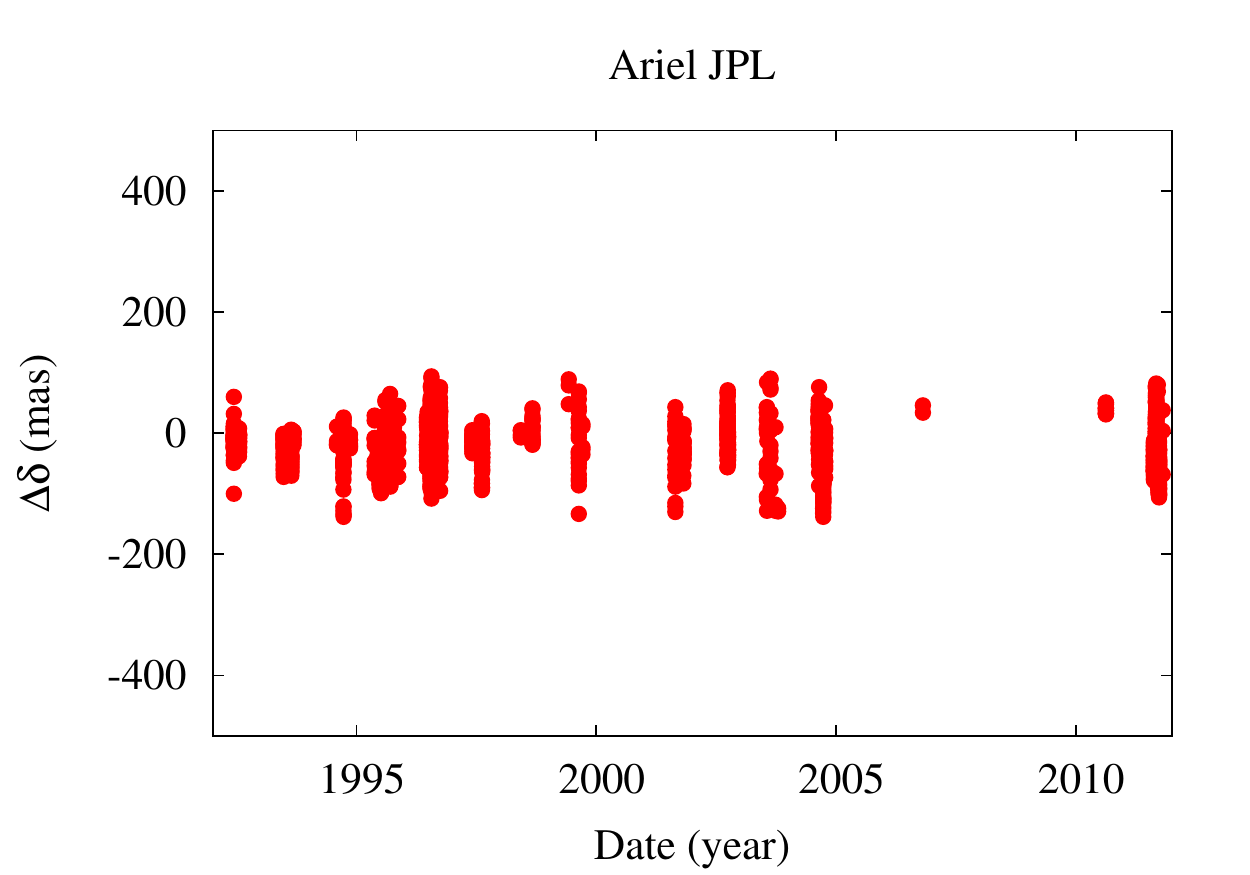}}
\caption{Distribution of the offsets for Ariel as a function of time.
             }
         \label{figure6}
   \end{figure}

\begin{figure}
   \centering{
   \includegraphics[width=6.5cm]{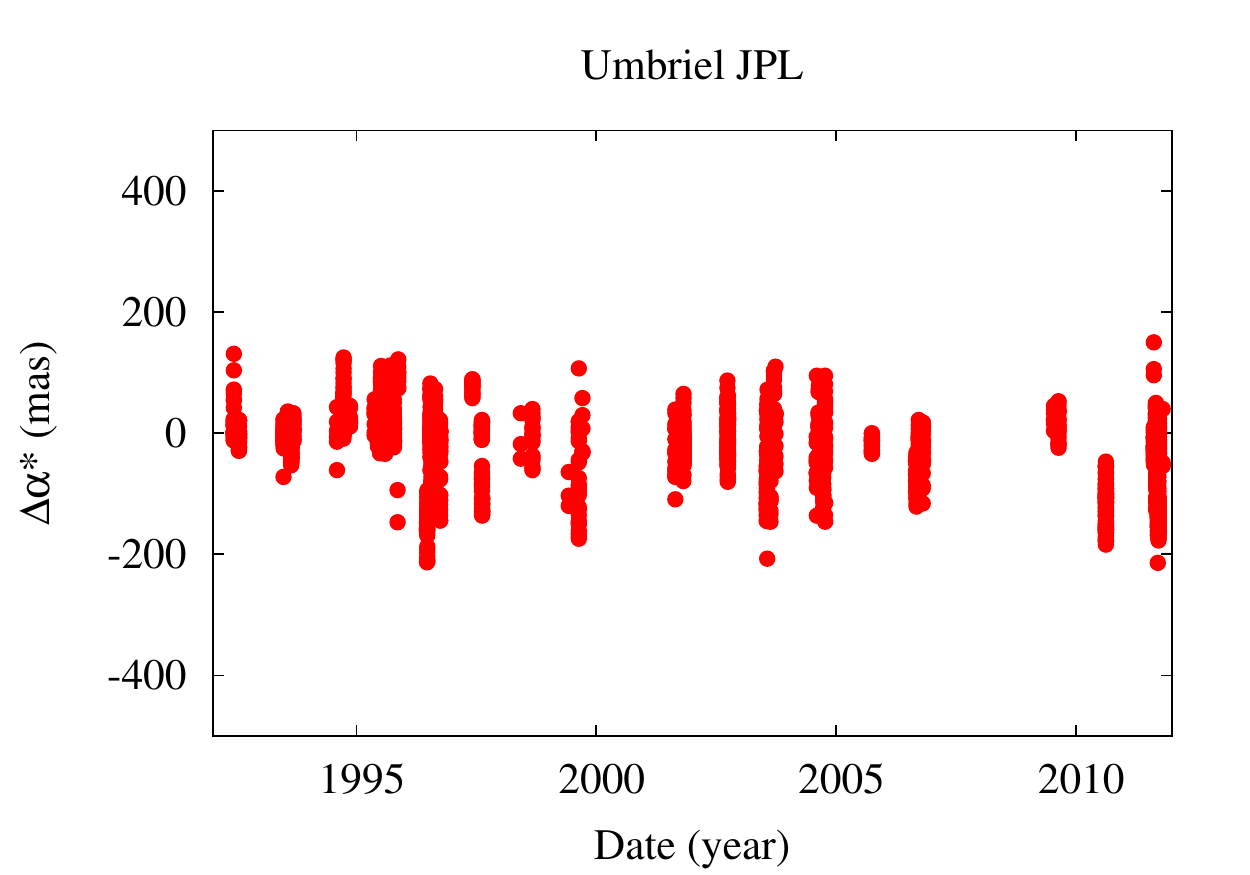}
   \includegraphics[width=6.5cm]{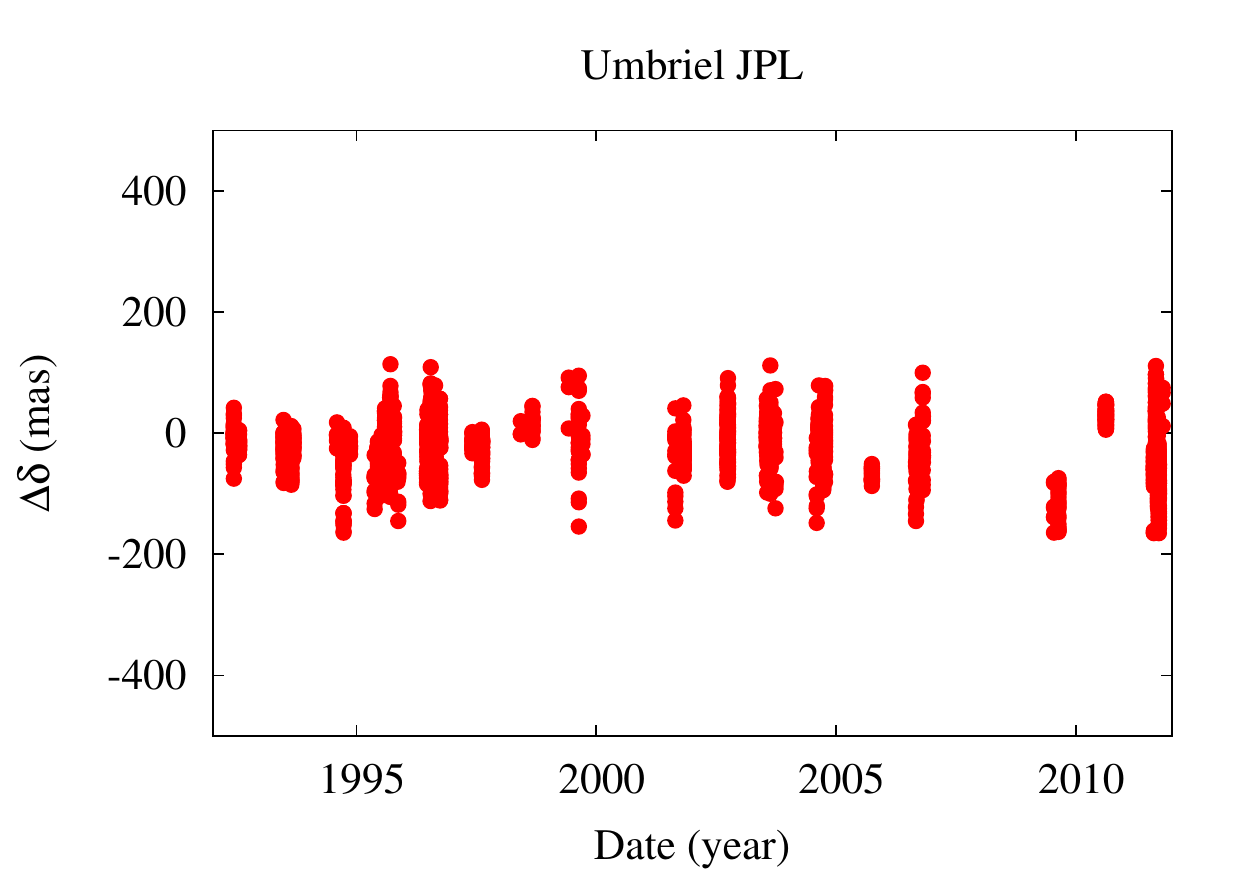}}
\caption{Distribution of the offsets for Umbriel as a function of time.
             }
         \label{figure7}
   \end{figure}

\begin{figure}
   \centering{
   \includegraphics[width=6.5cm]{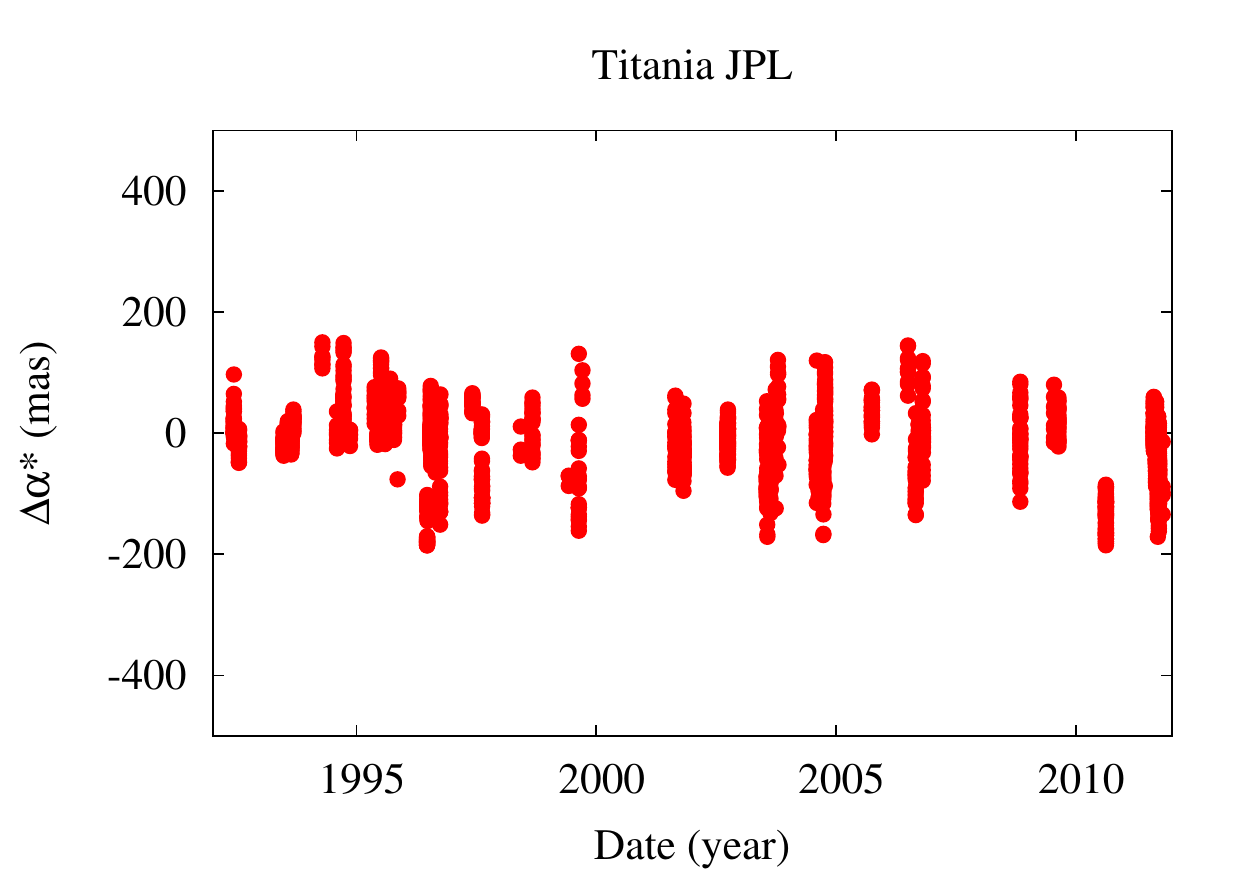}
   \includegraphics[width=6.5cm]{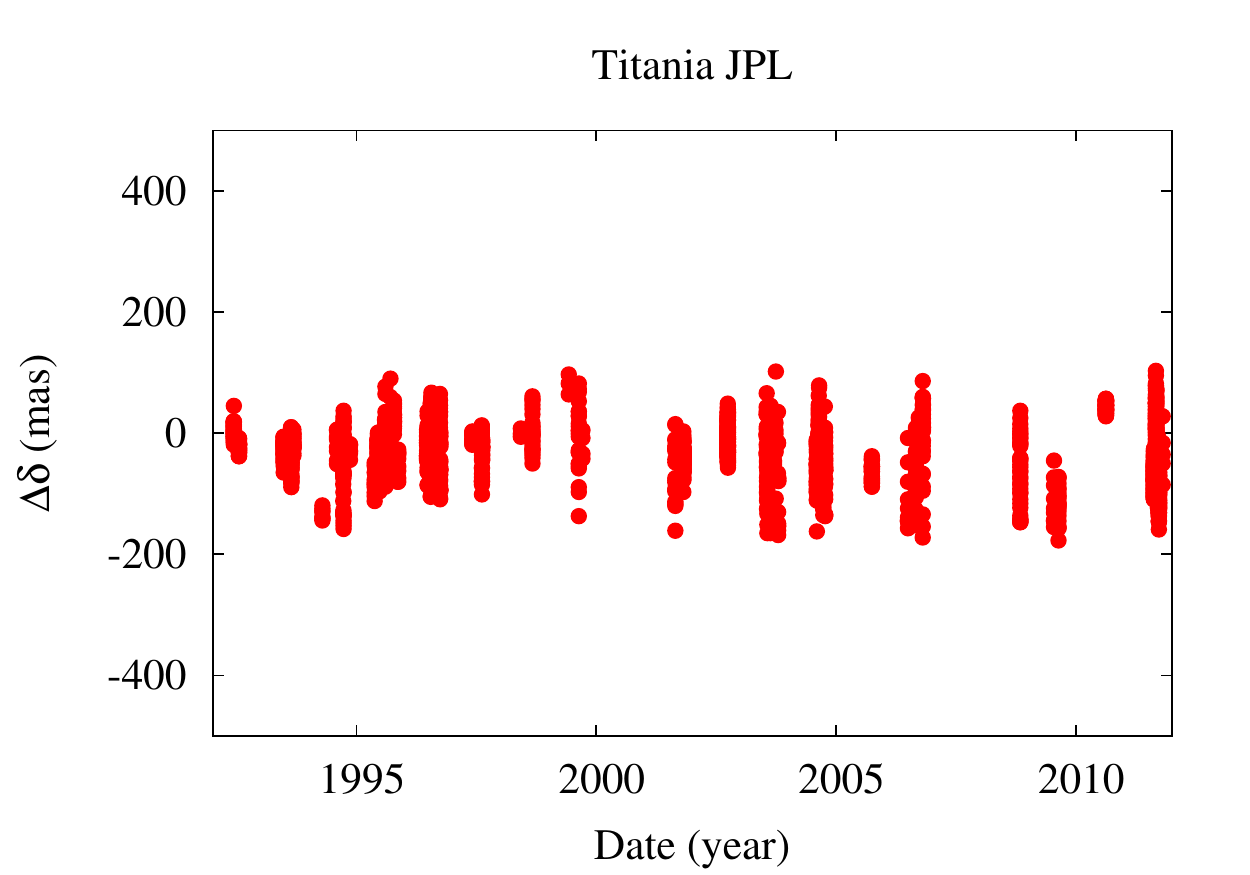}}
\caption{Distribution of the offsets for Titania as a function of time.
             }
         \label{figure8}
   \end{figure}

\begin{figure}
   \centering{
   \includegraphics[width=6.5cm]{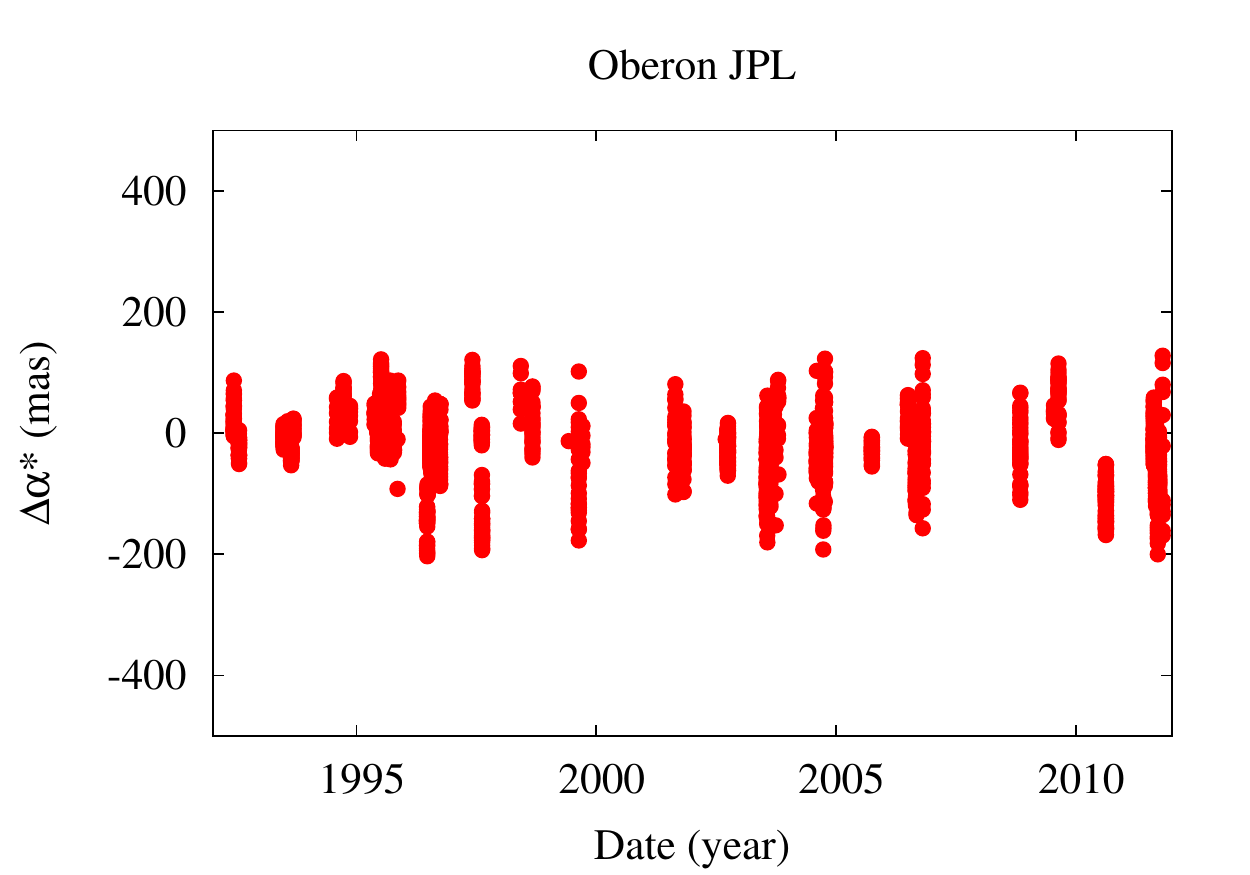}
   \includegraphics[width=6.5cm]{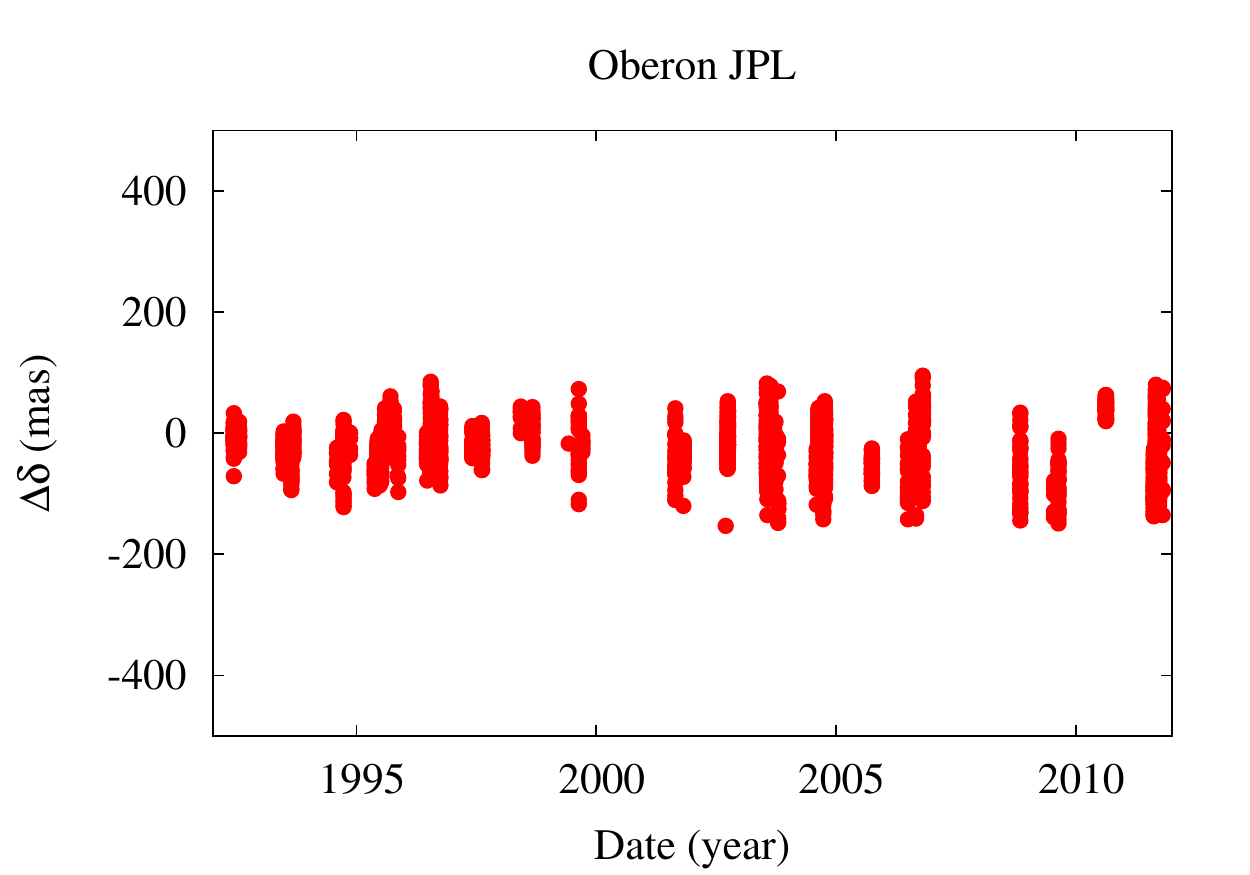}}
\caption{Distribution of the offsets for Oberon as a function of time.
             }
         \label{figure9}
   \end{figure}

Tables~\ref{table3} to \ref{table7} show yearly offsets for each satellite. Table~\ref{table3}
clearly shows the decrease in the number of positions of Miranda as the plane of its orbit approaches an edge-on view. 
Figures~\ref{figure5} to 
\ref{figure9} are provided to illustrate these tables by showing all final offsets for each satellite as 
a function of time. We use an asterisk to indicate
multiplication by the factor cos$\delta$ throughout the text.

\addtocounter{footnote}{1}\footnotetext{Generated by the Uranus Viewer Tool, PDS Rings Node, http://pds-rings.seti.org/\label{ft1}} 

\begin{table}
\caption{Results for Miranda}             
\label{table3}      
\begin{center}          
\begin{tabular}{c c c c c c c c}     
\hline\hline       
 Year & $\Delta\alpha*$ & $\Delta\delta$ & $\sigma_{\alpha}*$ & $\sigma_{\delta}$ & $e_{\alpha}*$ & $e_{\delta}$ &\#\\
  & 
\multicolumn{2}{c}{(mas)}& 
\multicolumn{2}{c}{(mas)}&
\multicolumn{2}{c}{(mas)}&
positions \\
\hline
1992 &  33  &  1   &   66  & 50  & 53  & 46 &    54\\
1993 & -32  & -22  &   88  & 60  & 48  & 54 &    71\\
1994 &  60  & -41  &   57  & 52  & 50  & 51 &    54\\
1995 &  16  & -2   &   56  & 62  & 45  & 48 &   158\\
1996 & -74  & -3   &   93  & 58  & 59  & 60 &   116\\
1997 & -89  & -22  &   87  & 40  & 62  & 44 &    36\\
1998 &  14  &  5   &   38  & 49  & 49  & 33 &    24\\
1999 & -116 &  35  &   62  & 59  & 44  & 51 &    22\\
2001 &  38  & -69  &   48  & 31  & 30  & 25 &    19\\
2002 & -63  &  23  &   149 & 58  & 53  & 53 &     6\\
2004 & -271 & -83  &   --  & --  & 52  & 34 &     1\\
2006 & -289 &  17  &   5   & 23  & 57  & 60 &     3\\
2009 & -103 &  25  &   --   & -- & 60  & 66 &     1\\
2011 & -164 &  38  &   107 & 61  & 47  & 49 &    19\\
\hline\hline                                    
\end{tabular}                                   
\end{center} 
Columns: year of observation; mean offset for the given year in right ascension; 
mean offset for the given year in declination; 
standard deviation of the offset in right ascension; standard deviation 
of the offset in declination; mean value of the standard deviations in right ascension
for the given year, as derived from the differences observed minus catalogue 
positions for the reference stars; 
mean value of the standard deviations in declination for the given year,
as derived from the differences observed minus catalogue positions for the
reference stars; number of positions that yielded the results shown in Cols. 
1 to 7. All angular units are in mas (milliarcsecond).
\end{table}

\begin{table}
\caption{Results for Ariel}             
\label{table4}      
\begin{center}          
\begin{tabular}{c c c c c c c c}     
\hline\hline       
 Year & $\Delta\alpha*$ & $\Delta\delta$ & $\sigma_{\alpha}*$ & $\sigma_{\delta}$ & $e_{\alpha}*$ & $e_{\delta}$ &\#\\
  & 
\multicolumn{2}{c}{(mas)}& 
\multicolumn{2}{c}{(mas)}&
\multicolumn{2}{c}{(mas)}&
positions \\
\hline
1992 &    8   & -13  &   34 &  20 &  54 &  48 &    75\\
1993 &   -12  & -28  &   19 &  20 &  49 &  53 &    95\\
1994 &    36  & -35  &   40 &  43 &  50 &  47 &    56\\
1995 &    28  & -23  &   43 &  38 &  50 &  50 &   206\\
1996 &   -40  & -4   &   74 &  43 &  54 &  51 &   245\\
1997 &   -37  & -25  &   89 &  27 &  56 &  45 &    78\\
1998 &   -4   &  4   &   37 &  17 &  50 &  36 &    28\\
1999 &   -69  & -8   &   77 &  53 &  48 &  51 &    33\\
2001 &   -14  & -30  &   39 &  38 &  44 &  43 &    46\\
2002 &    2   & -6   &   32 &  30 &  52 &  49 &    91\\
2003 &   -16  & -25  &   82 &  69 &  65 &  68 &    35\\
2004 &   -22  & -31  &   65 &  51 &  57 &  55 &   102\\
2006 &   -79  &  40  &   32 &  8  &  61 &  46 &     2\\
2010 &   -168 &  40  &   27 &  7  &  79 &  58 &    15\\
2011 &   -61  & -30  &   48 &  38 &  62 &  58 &   603\\
\hline\hline       
\end{tabular}
\end{center}
Columns: same as Table~\ref{table3}.                                    
\end{table}

\begin{table}
\caption{Results for Umbriel}             
\label{table5}      
\begin{center}          
\begin{tabular}{c c c c c c c c}     
\hline\hline       
 Year & $\Delta\alpha*$ & $\Delta\delta$ & $\sigma_{\alpha}*$ & $\sigma_{\delta}$ & $e_{\alpha}*$ & $e_{\delta}$ &\#\\
  & 
\multicolumn{2}{c}{(mas)}& 
\multicolumn{2}{c}{(mas)}&
\multicolumn{2}{c}{(mas)}&
positions \\
\hline
1992   &   8   &  -13  &   31 &  20 &  54 &  47  &      60\\
1993   &  -11  &  -30  &   21 &  25 &  49 &  54  &     115\\
1994   &   36  &  -54  &   37 &  48 &  51 &  49  &      68\\
1995   &   27  &  -24  &   39 &  45 &  49 &  49  &     230\\
1996   &  -36  &  -7   &   66 &  48 &  56 &  54  &     259\\
1997   &  -22  &  -23  &   77 &  20 &  56 &  47  &      68\\
1998   &  -8   &   15  &   32 &  15 &  52 &  36  &      23\\
1999   &  -63  &  -3   &   73 &  57 &  49 &  51  &      36\\
2001   &  -18  &  -33  &   36 &  35 &  36 &  37  &      60\\
2002   &  -9   &  -6   &   36 &  37 &  54 &  51  &     122\\
2003   &  -29  &  -13  &   72 &  43 &  62 &  58  &      91\\
2004   &  -31  &  -29  &   55 &  43 &  59 &  53  &     110\\
2005   &  -18  &  -69  &   11 &  10 &  43 &  54  &      19\\
2006   &  -45  &  -38  &   40 &  49 &  56 &  58  &      61\\
2009   &   17  &  -118 &   22 &  30 &  61 &  61  &      30\\
2010   &  -117 &   30  &   36 &  12 &  72 &  66  &      99\\
2011   &  -54  &  -47  &   56 &  48 &  57 &  52  &     536\\
\hline\hline       
\end{tabular}
\end{center} 
Columns: same as Table~\ref{table3}.                                    
\end{table}

\begin{table}
\caption{Results for Titania}             
\label{table6}      
\begin{center}          
\begin{tabular}{c c c c c c c c}     
\hline\hline       
 Year & $\Delta\alpha*$ & $\Delta\delta$ & $\sigma_{\alpha}*$ & $\sigma_{\delta}$ & $e_{\alpha}*$ & $e_{\delta}$ &\#\\
  & 
\multicolumn{2}{c}{(mas)}& 
\multicolumn{2}{c}{(mas)}&
\multicolumn{2}{c}{(mas)}&
positions \\
\hline
1992    &  -1   &  -10   &  27  &  17  &  54  &  48  &       69\\
1993    &  -8   &  -32   &  18  &  22  &  49  &  54  &      143\\
1994    &   43  &  -58   &  52  &  54  &  47  &  48  &       77\\
1995    &   34  &  -20   &  35  &  38  &  49  &  50  &      228\\
1996    &  -31  &  -15   &  67  &  43  &  55  &  52  &      285\\
1997    &  -22  &  -20   &  67  &  25  &  56  &  47  &       74\\
1998    &   3   &   1    &  33  &  30  &  49  &  35  &       29\\
1999    &  -57  &  -1    &  80  &  57  &  48  &  49  &       31\\
2001    &  -25  &  -47   &  34  &  34  &  39  &  40  &       65\\
2002    &  -11  &  -6    &  20  &  23  &  52  &  50  &      144\\
2003    &  -33  &  -41   &  61  &  53  &  65  &  60  &      118\\
2004    &  -28  &  -48   &  51  &  41  &  59  &  56  &      239\\
2005    &   38  &  -65   &  19  &  15  &  45  &  52  &       27\\
2006    &  -5   &  -39   &  70  &  64  &  56  &  58  &       82\\
2008    &  -10  &  -62   &  50  &  56  &  66  &  57  &       38\\
2009    &   19  &  -115  &  27  &  31  &  60  &  59  &       32\\
2010    &  -133 &   43   &  27  &  8   &  71  &  67  &       87\\
2011    &  -47  &  -51   &  46  &  45  &  59  &  56  &      820\\
\hline\hline       
\end{tabular}
\end{center} 
Columns: same as Table~\ref{table3}.                                    
\end{table}

\begin{table}
\caption{Results for Oberon}             
\label{table7}      
\begin{center}          
\begin{tabular}{c c c c c c c c}     
\hline\hline       
 Year & $\Delta\alpha*$ & $\Delta\delta$ & $\sigma_{\alpha}*$ & $\sigma_{\delta}$ & $e_{\alpha}*$ & $e_{\delta}$ &\#\\
  & 
\multicolumn{2}{c}{(mas)}& 
\multicolumn{2}{c}{(mas)}&
\multicolumn{2}{c}{(mas)}&
positions \\
\hline     
1992  &    2   & -8  &   31  & 16 & 54 & 48  &   77\\
1993  &   -15  & -33 &   20  & 24 & 48 & 54  & 119\\
1994  &    35  & -36 &   25  & 36 & 50 & 49  &   64\\
1995  &    24  & -15 &   40  & 34 & 49 & 49  & 230\\
1996  &   -36  & -9  &   56  & 33 & 53 & 49  &  327\\
1997  &   -31  & -21 &   100 & 17 & 56 & 48  &     93\\
1998  &    27  &  7  &   40  & 23 & 49 & 36  &    36\\
1999  &   -55  & -18 &   66  & 42 & 48 & 47  &    36\\
2001  &   -21  & -39 &   41  & 31 & 41 & 43  &  65\\
2002  &   -25  & -9  &   19  & 29 & 53 & 51  &  154\\
2003  &   -39  & -17 &   56  & 47 & 63 & 61  &  163\\
2004  &   -32  & -34 &   43  & 41 & 58 & 56  &  295\\
2005  &   -29  & -56 &   13  & 17 & 45 & 53  &     33\\
2006  &   -32  & -32 &   58  & 54 & 55 & 58  &   142\\
2008  &   -21  & -60 &   46  & 50 & 67 & 58  &   39\\
2009  &    55  & -88 &   33  & 38 & 60 & 59  & 39\\
2010  &   -110 &  42 &   31  & 12 & 71 & 68  &   101\\
2011  &   -62  & -41 &   44  & 41 & 57 & 55  &    915\\
\hline\hline       
\end{tabular}
\end{center}
Columns: same as Table~\ref{table3}.                                    
\end{table}

The standard deviations, as shown in Cols. 4 and 5 in Tables~\ref{table3} 
to \ref{table7}, peak in right ascension. Although it is not easy to clearly identify the 
reason for this feature, some hypothesis may be investigated. 

Known small mechanical problems that affect the telescope tracking system might be 
responsible for a degradation in the quality of the measurements in right ascension. 
However, in this case, standard deviations in right ascension
should be systematically larger than those in declination, and this is not verified.

\begin{figure}
   \centering{
   \includegraphics[width=9cm]{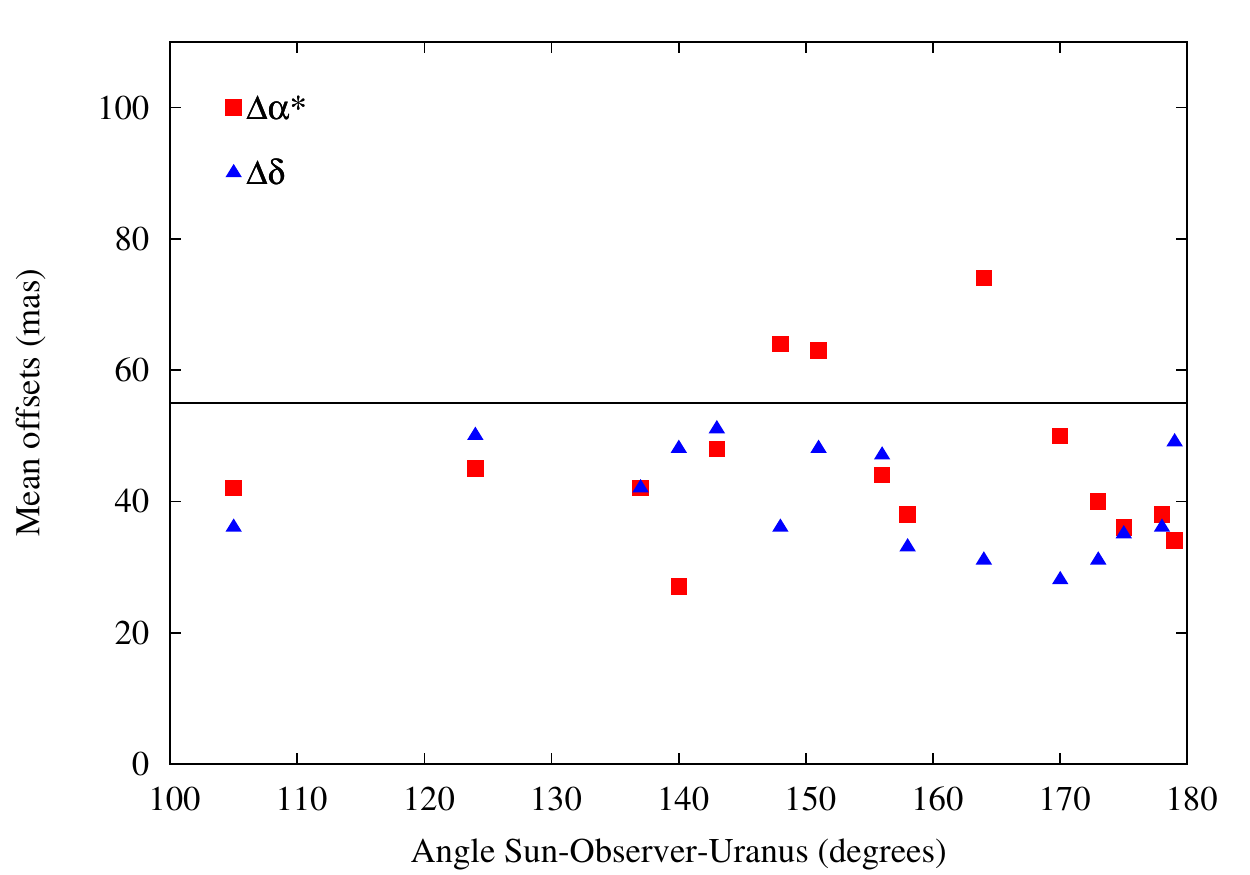}}
\caption{Offsets in right ascension (squares) and declination (triangles) as a function of the angle Sun-observer-Oberon. 
Each symbol is the average of 200 offsets. The horizontal line was set at ordinate 55 mas. Only data from
Oberon were used here.
             }
         \label{figure10}
   \end{figure}

Another source of larger uncertainties are occurrences of incorrect timing inserted 
in the image headers. This mostly affects the right ascension measurements and is known
to have happened in the past (before 2000). Figure~\ref{figure10} illustrates this point.
We note that both mechanical and timing problems may slightly increase or decrease the 
values in right ascension in our case. We therefore expect on average larger standard 
deviations and not a systematic effect.

The only points above the threshold of 55 mas in Fig.~\ref{figure10} are those
in right ascension and not far from opposition, where the movement of Uranus (most of it along right
ascension) is faster. The value of 55 mas is arbitrary to some extent and was chosen because the 
standard deviations in declination (Col. 5 in Tables~\ref{table3} to \ref{table7}) are almost 
all smaller than it. Therefore, some timing problem is a plausible and probable explanation. Figure~\ref{figure10}
also shows a number of observations far from opposition that are therefore helpful for determining the 
heliocentric distance of Uranus and its satellites.

\begin{figure}
   \centering{
   \includegraphics[width=9cm]{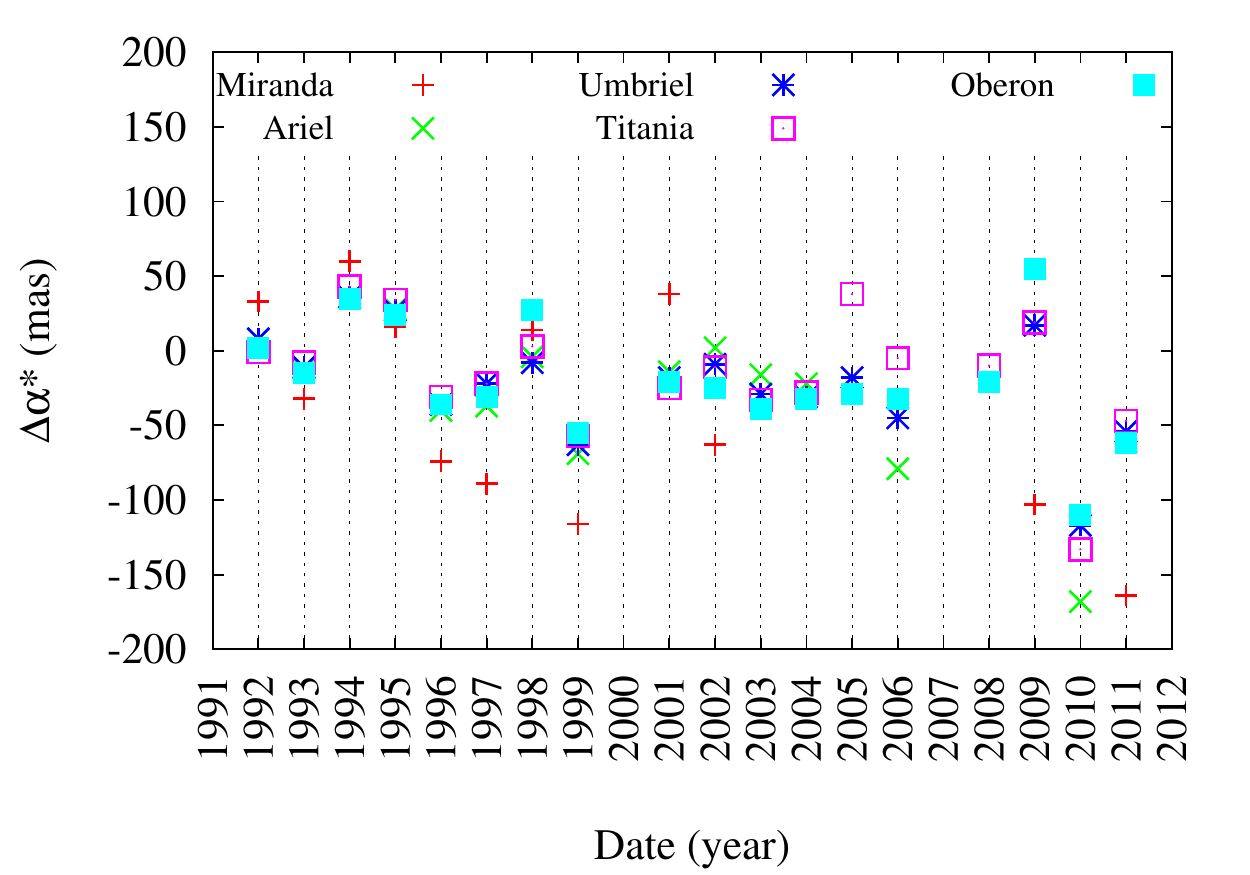}
   \includegraphics[width=9cm]{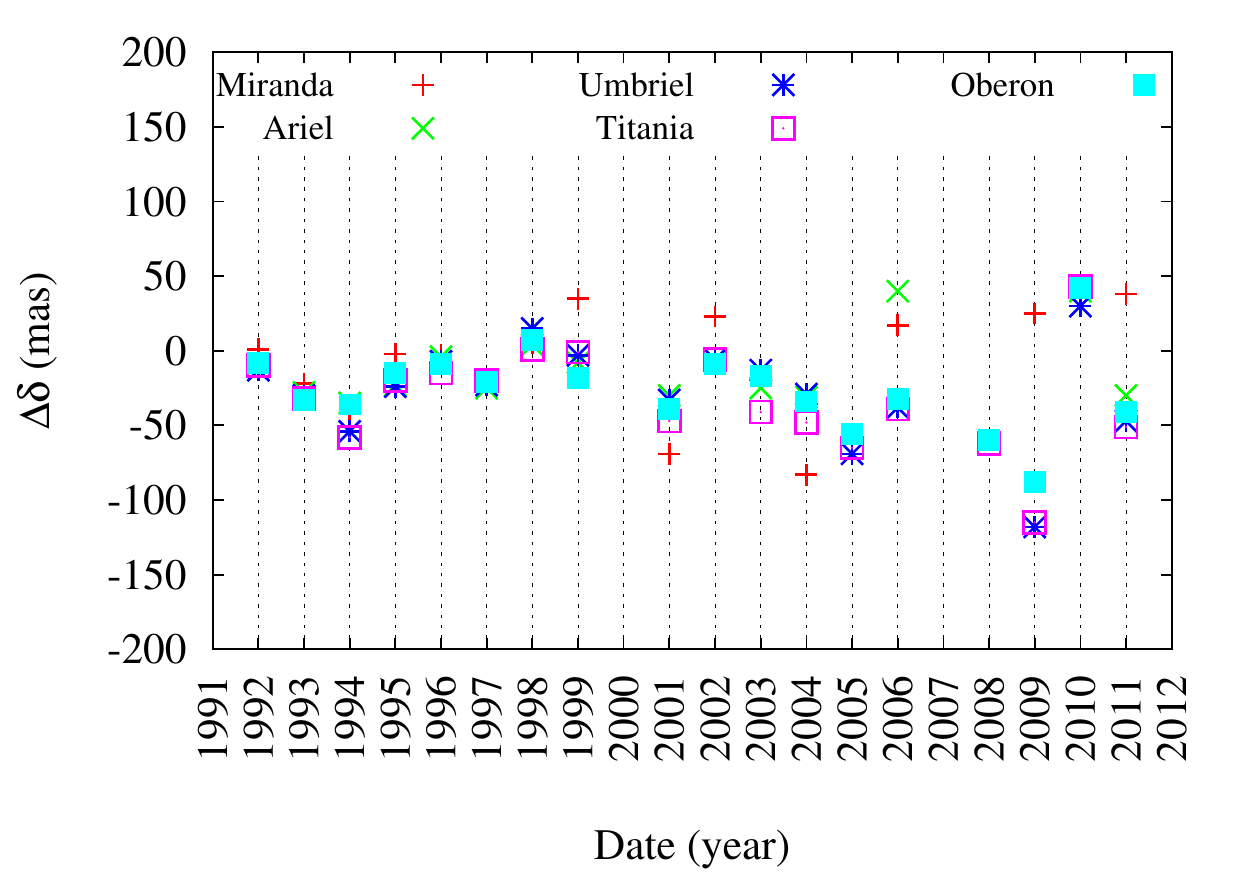}}
\caption{Offsets in right ascension (upper panel) and declination (lower panel), as given by
Tables~\ref{table3} to \ref{table7}, to all five satellites. Note that the offsets in right ascension
for Miranda, measured in 2004 and 2006, fall outside the limits of the {\it y}-axis.
             }
         \label{figure11}
   \end{figure}

Another interesting point about the offsets is shown in Fig.~\ref{figure11} where it is possible to note that,
except for Miranda, the yearly offsets in right ascension and declination are similiar. The overall 
offsets for all five satellites are shown in Table~\ref{table8}. 
This is an indication that the offsets we find are mostly due to the planetary ephemeris DE432, that is,
due to the ephemeris position of the barycenter of the Uranus system. 
Support for this result is also obtained from Cols. 2 and 3 in Table~\ref{table9}.

\begin{table}
\caption{Overall offsets, all satellites}             
\label{table8}      
\begin{center}          
\begin{tabular}{l c c c c c c}     
\hline\hline       
Satellite & $\Delta\alpha*$ & $\Delta\delta$ & $\sigma_{\alpha}*$ & $\sigma_{\delta}$ &\#\\
  & 
\multicolumn{2}{c}{(mas)}& 
\multicolumn{2}{c}{(mas)}&
positions \\
\hline     
Miranda  & -22 & -8 & 96 & 60 & 584\\
Ariel & -30 & -21 & 65 & 40 & 1710\\
Umbriel & -28 & -27 & 62 & 48 & 1987\\
Titania & -25 & -35 & 59 & 48 & 2588\\
Oberon & -35 & -26 & 56 & 42 & 2928\\
\hline\hline       
\end{tabular}
\end{center}
Columns: see Table~\ref{table3}.
\end{table}

\begin{table}
\caption{Overall residuals with respect to Oberon}             
\label{table9}      
\begin{center}          
\begin{tabular}{l c c c c c c}     
\hline\hline       
Satellite & $[\Delta\alpha*]$ & $[\Delta\delta]$ & $\sigma_{\alpha}*$ & $\sigma_{\delta}$ &\#\\
  & 
\multicolumn{2}{c}{(mas)}& 
\multicolumn{2}{c}{(mas)}&
positions \\
\hline     
Miranda  & -2 & 7 & 69 & 54 & 584\\
Ariel & 4 & 1 & 31 & 28 & 1710\\
Umbriel & 3 & -6 & 33 & 30 & 1987\\
Titania & 5 & -6 & 31 & 30 & 2588\\
\hline\hline       
\end{tabular}
\end{center}
Columns 2 and 3 give the difference in the sense satellite
offset minus Oberon offset. Columns 4 and 5 give the
respective standard deviations of these offsets.
\end{table}

\begin{figure}
   \centering{
   \includegraphics[width=6.5cm]{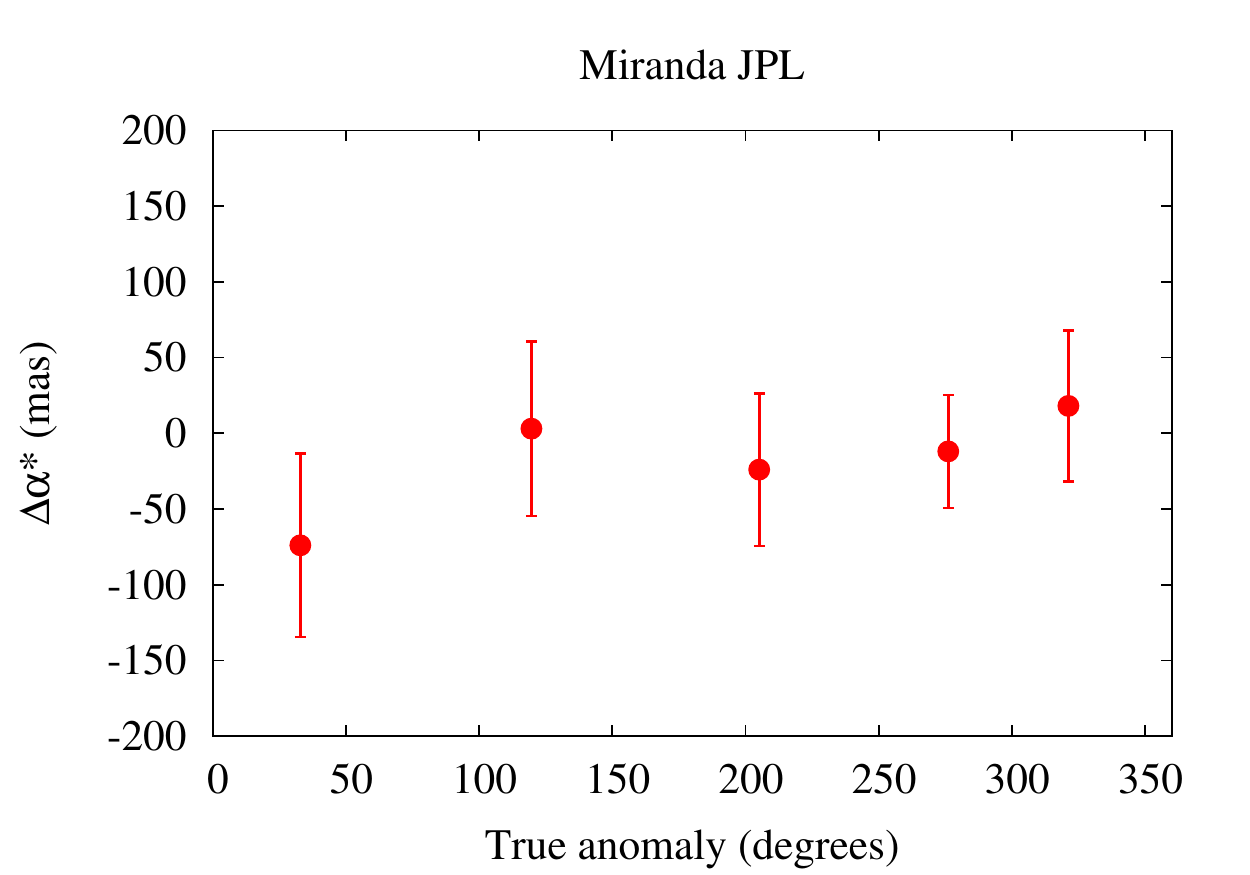}
   \includegraphics[width=6.5cm]{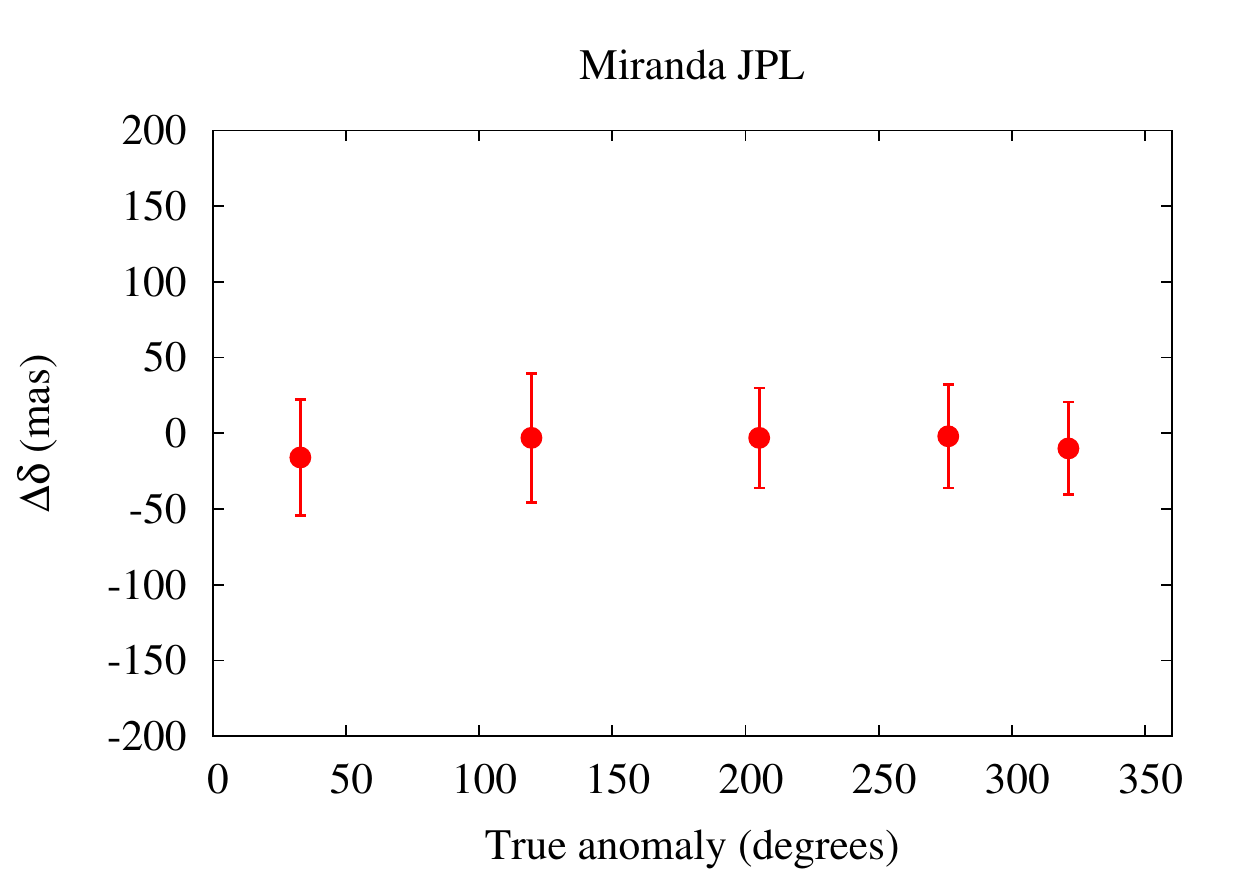}}      
\caption{Distribution of the offsets for Miranda as a function of the true anomaly. Solid dots:
orbit view close to face-on. Note that there were not enough positions of Miranda with
an orbit view close to the edge-on to appear in these plots.
             }
         \label{figure12}
   \end{figure}

\begin{figure}
   \centering{
   \includegraphics[width=6.5cm]{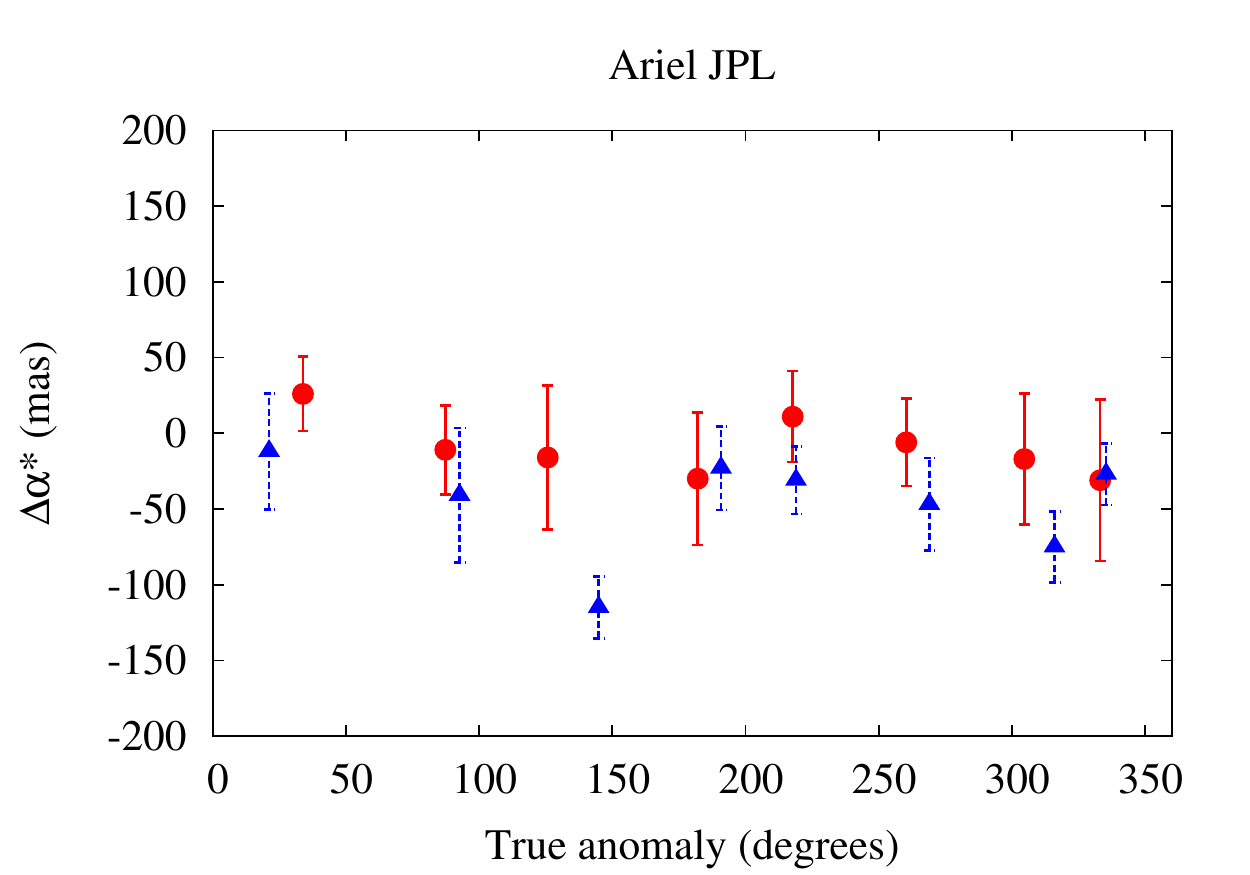}
   \includegraphics[width=6.5cm]{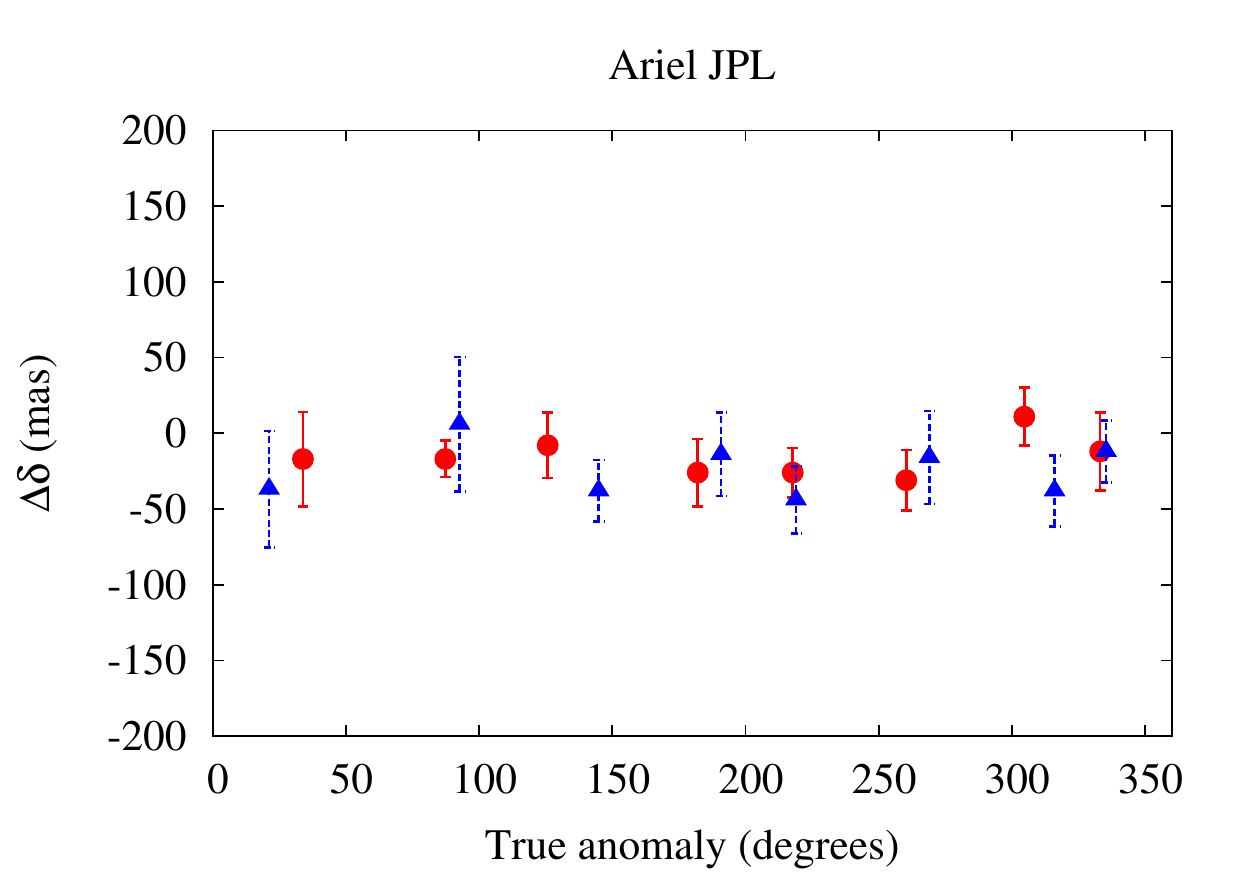}}      
\caption{Distribution of the offsets for Ariel as a function of the true anomaly. Solid dots:
orbit view close to face-on. Solid triangles: orbit view close to edge-on.
             }
         \label{figure13}
   \end{figure}

\begin{figure}
   \centering{
   \includegraphics[width=6.5cm]{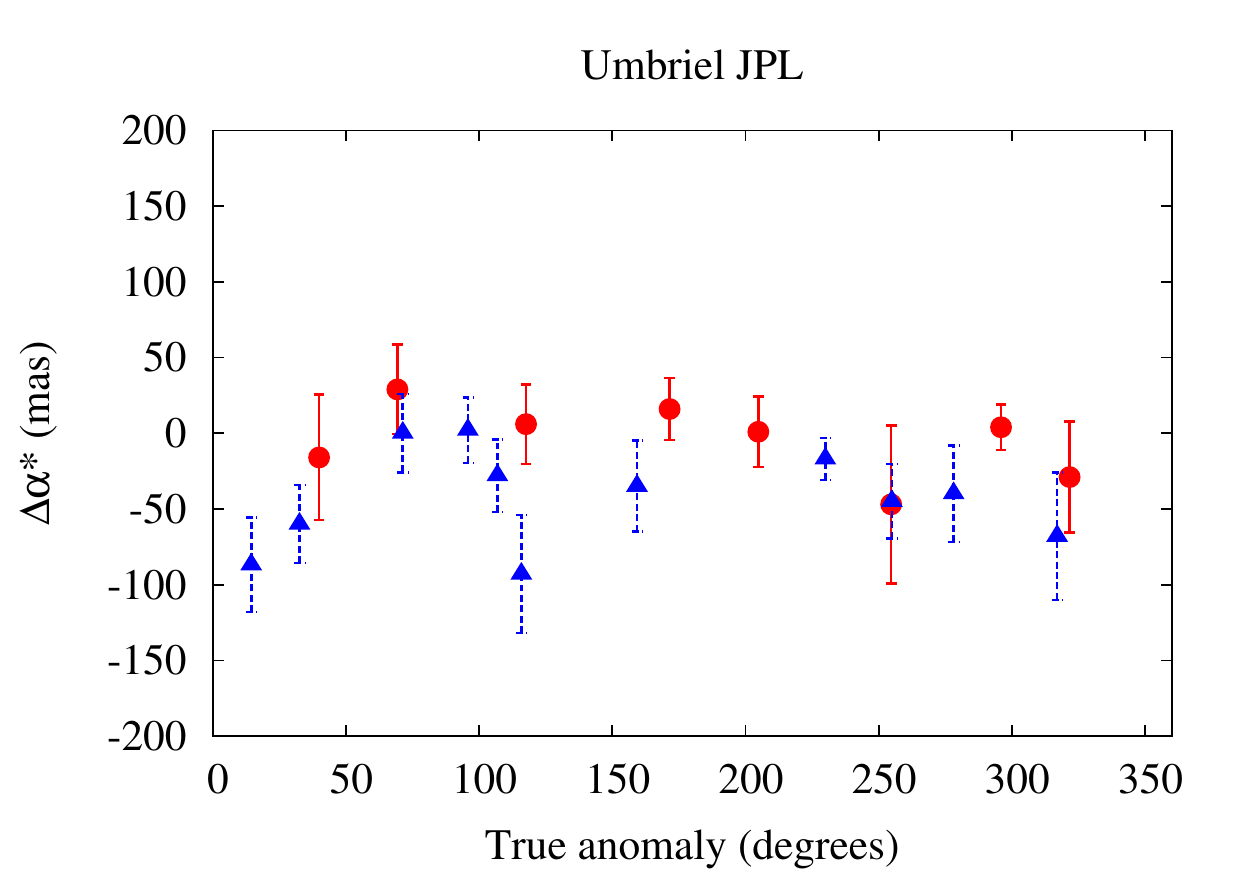}
   \includegraphics[width=6.5cm]{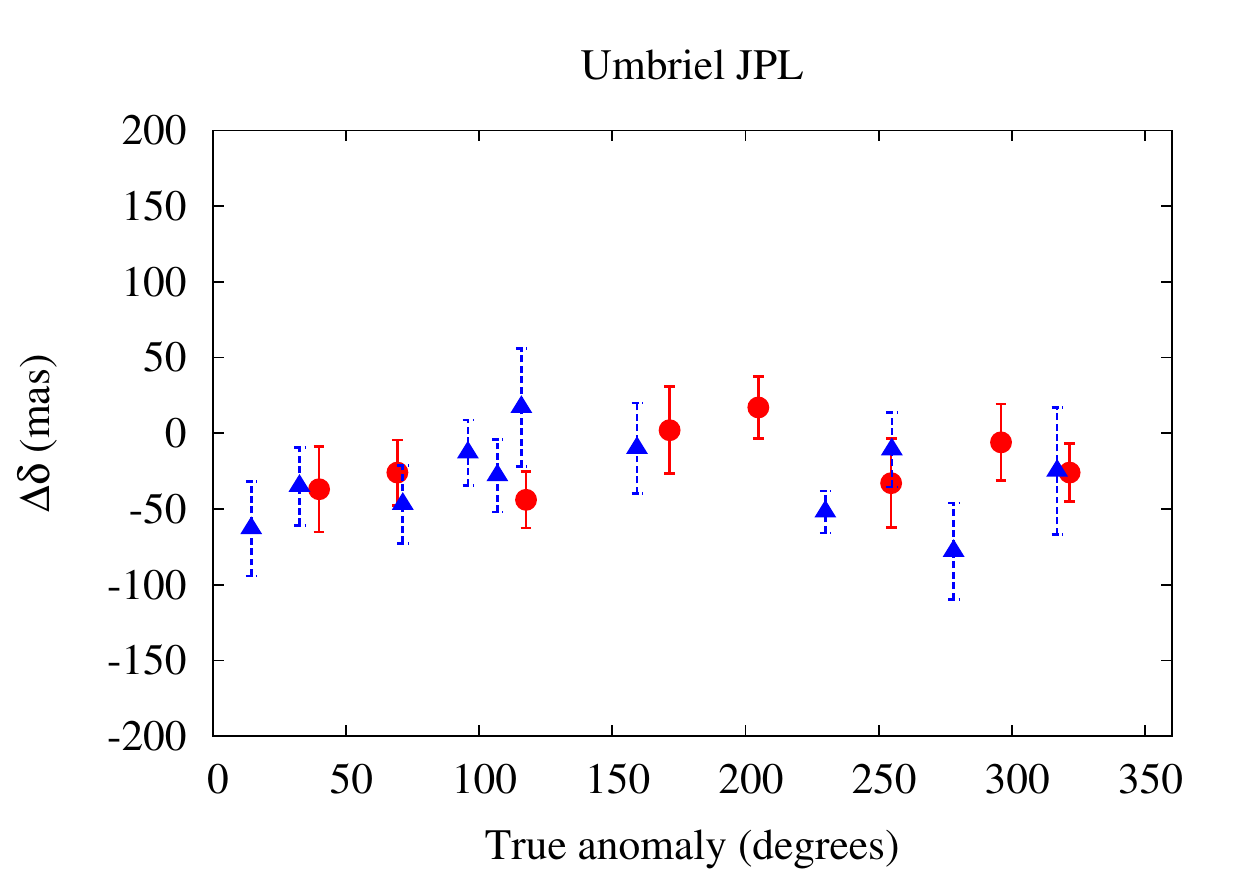}}      
\caption{Distribution of the offsets for Umbriel as a function of the true anomaly. Solid dots:
orbit view close to face-on. Solid triangles: orbit view close to edge-on.
             }
         \label{figure14}
   \end{figure}

\begin{figure}
   \centering{
   \includegraphics[width=6.5cm]{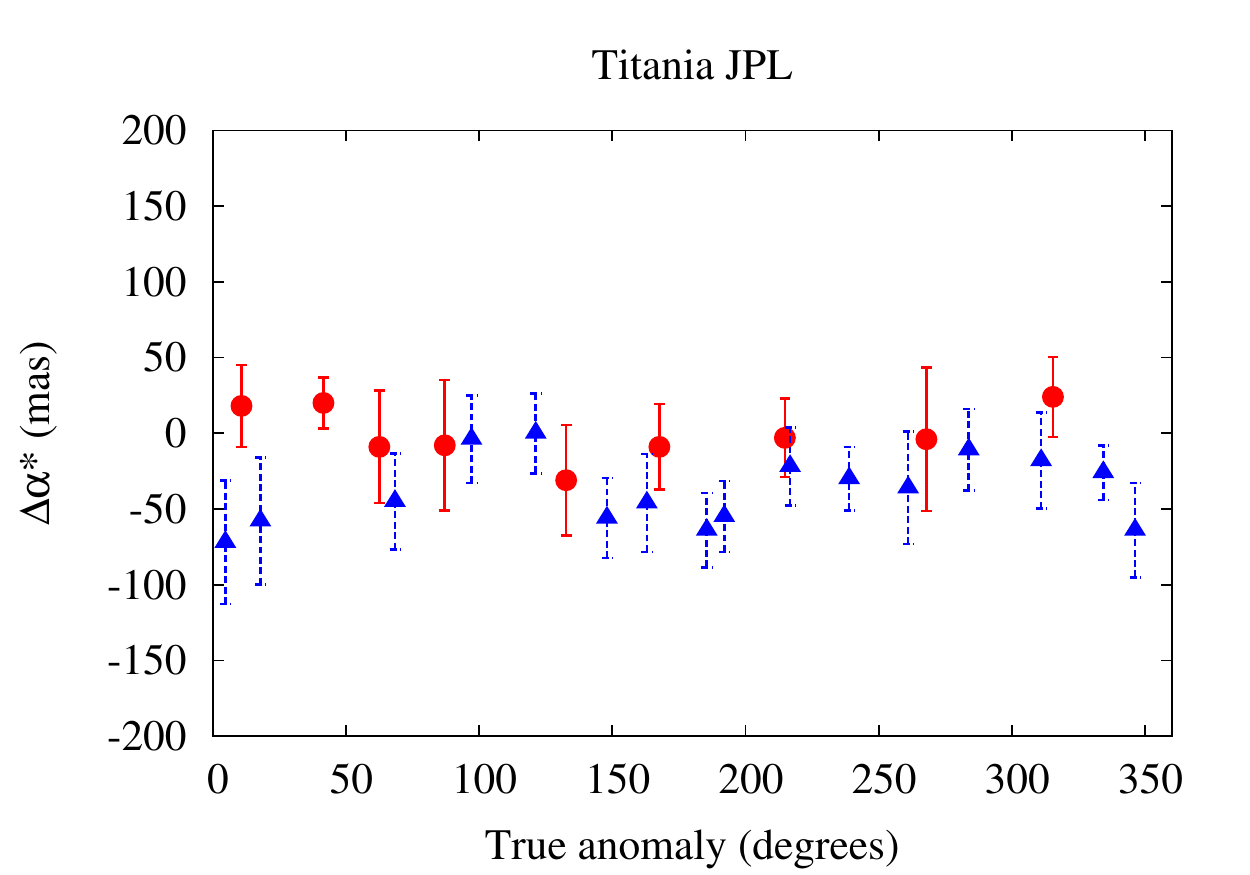}
   \includegraphics[width=6.5cm]{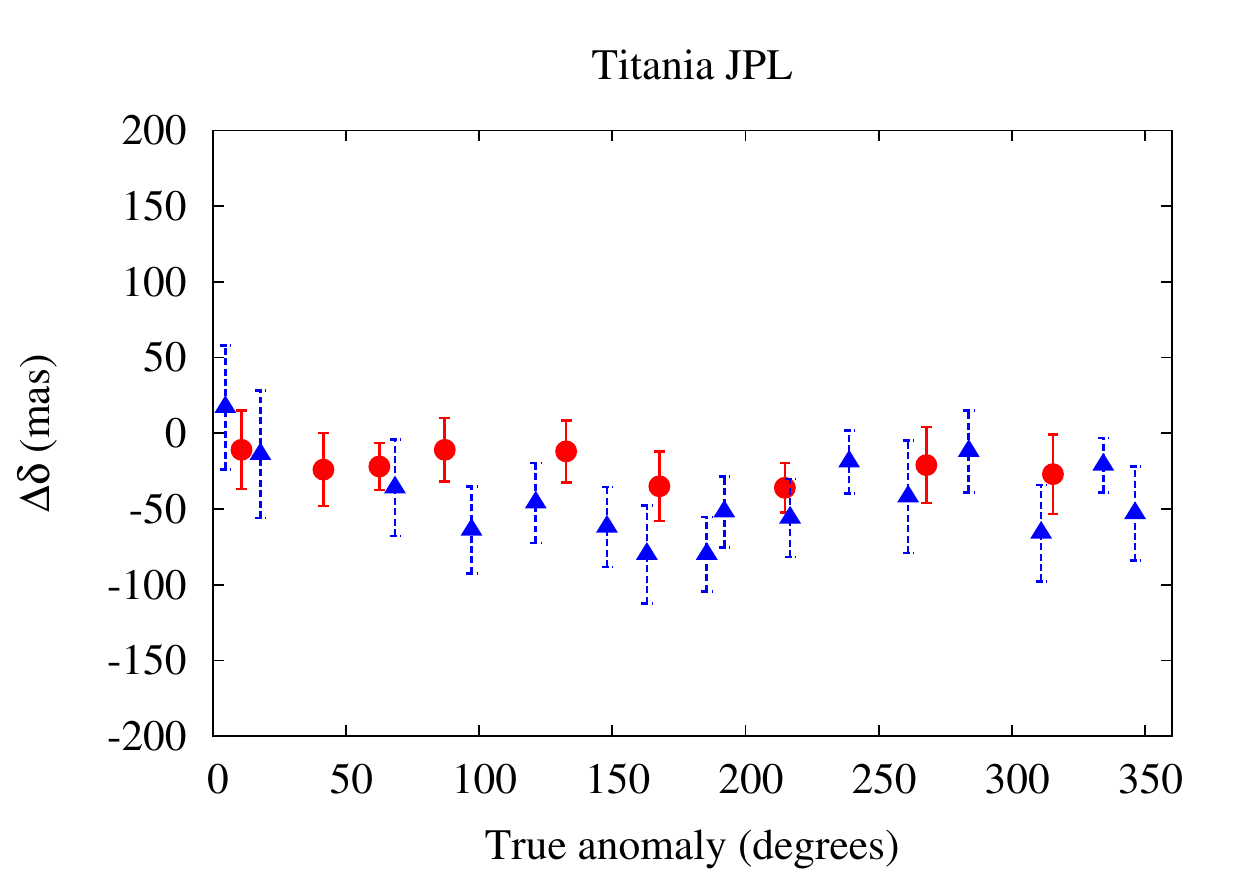}}
\caption{Distribution of the offsets for Titania as a function of the true anomaly. Solid dots:
orbit view close to face-on. Solid triangles: orbit view close to edge-on.
             }
         \label{figure15}
   \end{figure}

\begin{figure}
   \centering{
   \includegraphics[width=6.5cm]{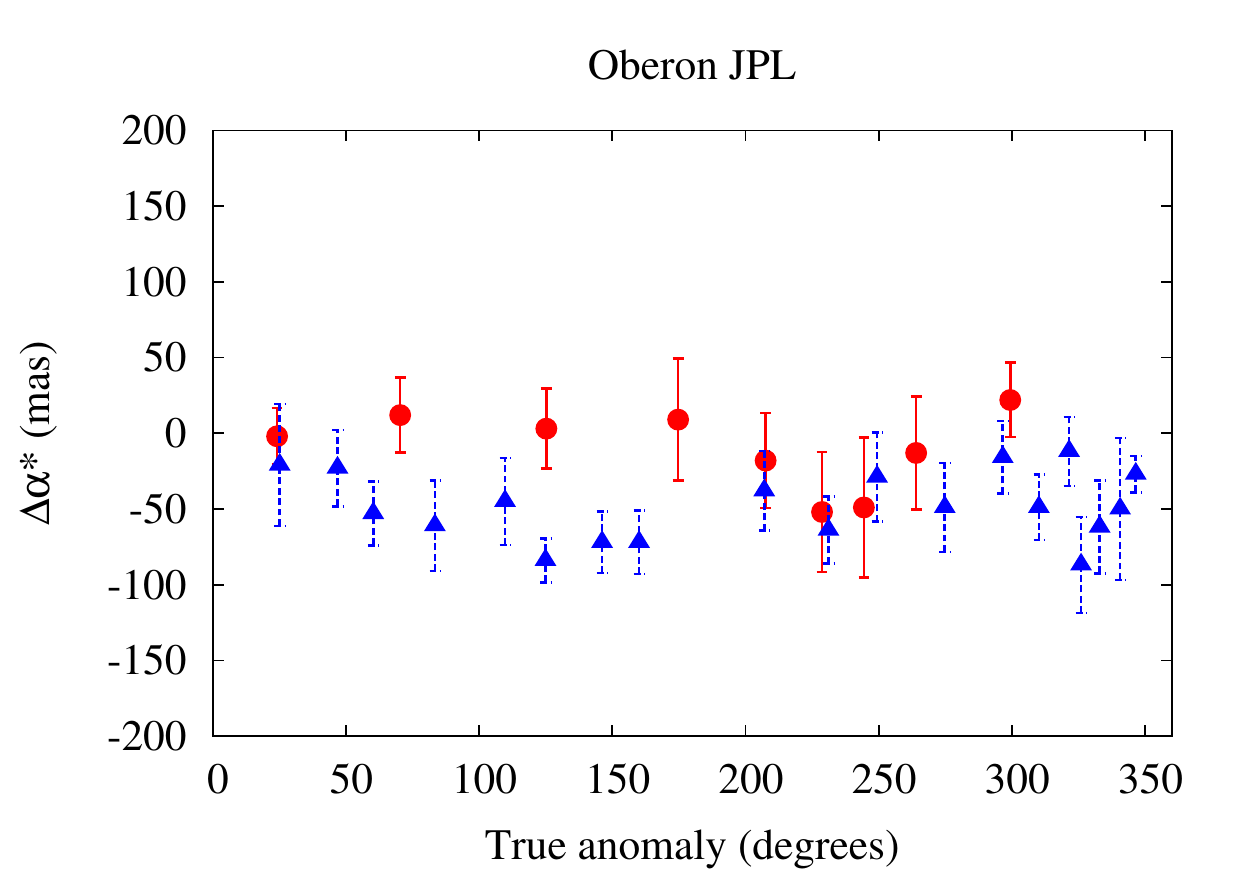}
   \includegraphics[width=6.5cm]{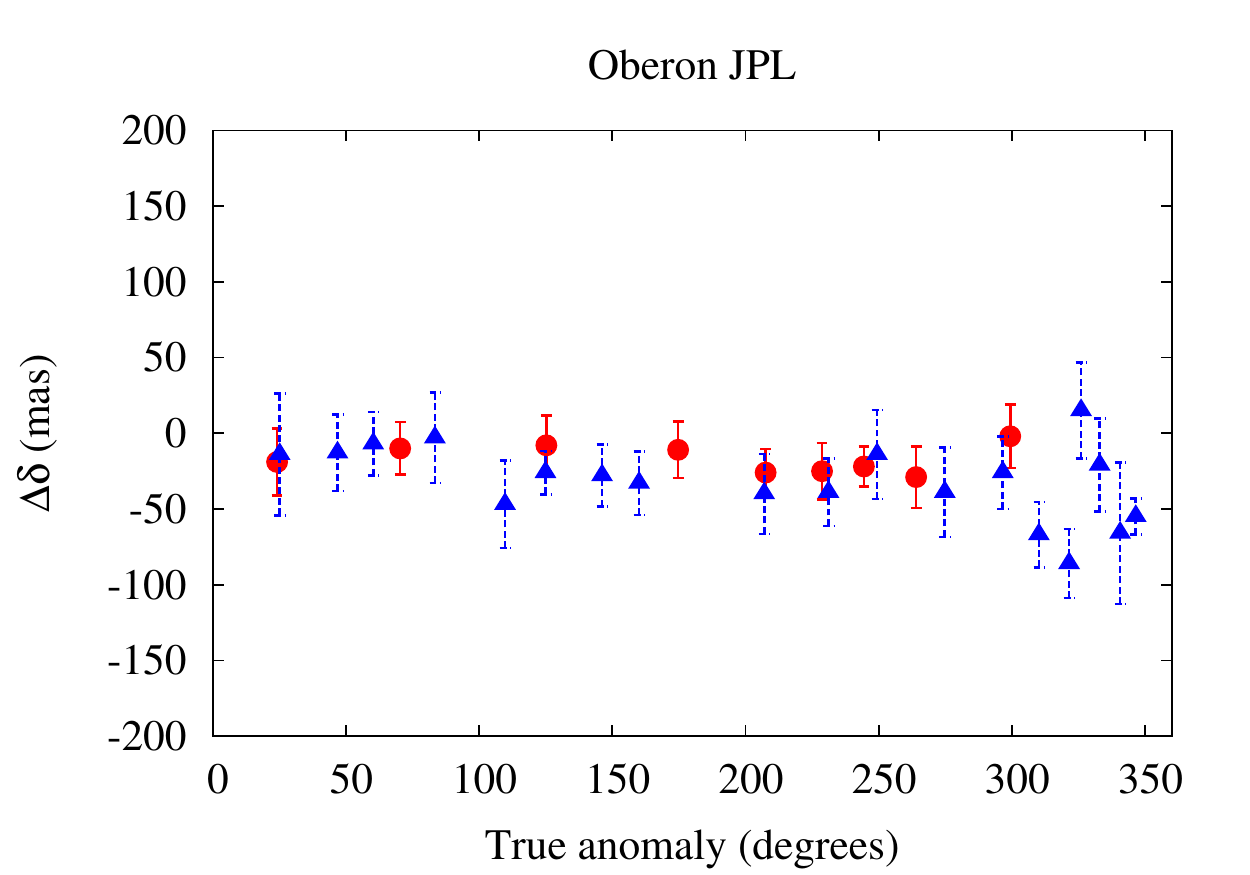}}      
\caption{Distribution of the offsets for Oberon as a function of the true anomaly. Solid dots:
orbit view close to face-on. Solid triangles: orbit view close to edge-on.
             }
         \label{figure16}
   \end{figure}

At this point, it is interesting to check whether our observed positions might also be an improvement
to models of the satellite motions. A simple way of checking this is given by Figs.~\ref{figure12}
to \ref{figure16}. In them, each symbol is the mean of 100 offsets, and the length of the error bars,
above and below the respective dots or triangles, is three times the error of the mean.

It is possible to note systematic deviations in the offsets of both right ascension and declination
as a function of the true anomaly. Positions from observations made with the satellite orbits close to edge-on
tend to be more affected by systematic effects than those from observations made with the satellite orbits close to face-on. However, both sets of positions present systematic effects, and a theory that corrects 
for them would mean an improvement to the dynamical models for the motions of these satellites.

\section{Comparisons with other ephemerides}

We also compared our positions with those obtained from the recent planetary ephemeris 
INPOP13c\footnote{See http://www.imcce.fr/inpop/ for the ephemeris and the respective scientific notes.}
and from the dynamical model for the motion of the five main satellites of Uranus
NOE-7-2013\footnote{ftp.imcce.fr/pub/ephem/satel/NOE/URANUS/SPICE/}. This model is a more
recent adjustment of the motions of the satellites than that given by
\citetads{2008P&SS...56.1766L}.

In this section, only observed positions of Oberon are used for the comparisons.

\subsection{INPOP13c}

Table~\ref{table10} lists the yearly offsets in the sense observed minus ephemeris (INPOP13c$+$ura111)
positions. We note that the right ascension between INPOP13c and our
observations agrees better than in DE432. However, the opposite is observed for the offsets in declination.
The very last line of Table~\ref{table10} shows information that is equivalent to that in 
Table~\ref{table8} for Oberon, according to INPOP13c$+$ura111, and can be 
assumed as a proxy to the offsets of the other satellites.

\begin{table}
\caption{Results for Oberon with INPOP13c$+$ura111}             
\label{table10}      
\begin{center}          
\begin{tabular}{c c c c c c c c}     
\hline\hline       
 Year & $\Delta\alpha*$ & $\Delta\delta$ & $\sigma_{\alpha}*$ & $\sigma_{\delta}$ & $e_{\alpha}*$ & $e_{\delta}$ &\#\\
  & 
\multicolumn{2}{c}{(mas)}& 
\multicolumn{2}{c}{(mas)}&
\multicolumn{2}{c}{(mas)}&
positions \\
\hline     
1992  &  -6  & -45 &   31  & 16 & 54 & 48  &    77\\
1993  & -20 & -70 &   20  & 24 & 48 & 54  & 119\\
1994  &   35 & -72 &   25  & 37 & 50 & 49  &    64\\
1995  &   28 & -51 &   40  & 34 & 49 & 49  &  230\\
1996  & -28 & -43 &   56  & 33 & 53 & 49  & 327\\
1997  & -19 & -54 &   100 & 17 & 56 & 48  &  93\\
1998  &   43 & -24 &   40  & 23 & 49 & 36  &    36\\
1999  & -35 & -47 &   66  & 42 & 48 & 47  &   36\\
2001  &    5  & -63 &   41  & 31 & 41 & 43  &     65\\
2002  &    4  & -30 &   19  & 29 & 53 & 51  &   154\\
2003  &  -7  & -35 &   56  & 47 & 63 & 61  &  163\\
2004  &   2  & -49 &   43  & 41 & 58 & 56  &   295\\
2005  &   8  & -68 &   13  & 17 & 45 & 53  &     33\\
2006  &   6  & -41 &   57  & 54 & 55 & 58  &   142\\
2008  &  19 & -63 &   46  & 50 & 67 & 58  &    39\\
2009  &  95 & -89 &   33  & 38 & 60 & 59  &    39\\
2010  & -69 &  44 &   31  & 12 & 71 & 68  &  101\\
2011  & -20 & -37 &   44  & 41 & 57 & 55  & 915\\
\hline
All years & -8 & -42 & 52 & 42 & -- & -- & 2928 \\
\hline\hline       
\end{tabular}
\end{center}
Columns: same as Table~\ref{table3}.                                    
\end{table}

\subsection{NOE-7-2013}

Table~\ref{table11} gives the yearly offsets in the sense observed minus ephemeris 
(DE432$+$NOE-7-2-13) positions. The results are similar to those given
in Table~\ref{table7}. This is expected since, as mentioned earlier in the text, the main source of
differences between observed and ephemeris positions is the planetary ephemerides,
which are responsible for the description of the motion of the barycenter of the Uranus system around the 
barycenter of the solar system.

\begin{table}
\caption{Results for Oberon with DE432$+$NOE-7-2013}             
\label{table11}      
\begin{center}          
\begin{tabular}{c c c c c c c c}     
\hline\hline       
 Year & $\Delta\alpha*$ & $\Delta\delta$ & $\sigma_{\alpha}*$ & $\sigma_{\delta}$ & $e_{\alpha}*$ & $e_{\delta}$ &\#\\
  & 
\multicolumn{2}{c}{(mas)}& 
\multicolumn{2}{c}{(mas)}&
\multicolumn{2}{c}{(mas)}&
positions \\
\hline     
1992  & -2  & -9 &   31  & 16 & 54 & 48  &   77\\
1993  & -13 & -34 &   20  & 24 & 48 & 54  & 119\\
1994  &  35 & -37 &   25  & 36 & 50 & 49  &   64\\
1995  &  26 & -16 &   40  & 34 & 49 & 49  & 230\\
1996  & -34 & -9 &   54  & 33 & 53 & 49  &  327\\
1997  & -27 & -22 &   97 & 17 & 56 & 48  &     93\\
1998  &  27 & 7 &   42  & 23 & 49 & 36  &    36\\
1999  & -51 & -17 &   66  & 42 & 48 & 47  &    36\\
2001  &  -19  & -40 &   40  & 30 & 41 & 43  &  65\\
2002  &  -19  & -9 &   18  & 28 & 53 & 51  &  154\\
2003  & -39  & -19 &   55  & 47 & 63 & 61  &  163\\
2004  &  -30 & -34 &   45  & 41 & 58 & 56  &  295\\
2005  &  -33  & -56 &   13  & 17 & 45 & 53  &     33\\
2006  &  -33  & -33 &   59  & 53 & 55 & 58  &   142\\
2008  &  -28 & -61 &   46  & 50 & 67 & 58  &   39\\
2009  &  55 & -90 &   34  & 37 & 60 & 59  & 39\\
2010  & -103 &  44 &   31  & 12 & 71 & 68  &   101\\
2011  & -63 & -42 &   45  & 40 & 57 & 55  &    915\\
\hline
All years & -34 & -27 & 57 & 42 & -- & -- & 2928 \\
\hline\hline       
\end{tabular}
\end{center}
Columns: same as Table~\ref{table3}.                                    
\end{table}

When we compare the offsets presented in Tables~\ref{table10} and \ref{table7} with
those presented in Tables~\ref{table11} and \ref{table7}, we note that the planetary
ephemerides clearly agree less well between each other than the models for the satellite motions (at least for the bodies studied here).

\section{Previous data from the Pico dos Dias Observatory for the main satellites of Uranus}

Raw images of the Uranus system, made at the Pico dos Dias observatory 
before 1992, are no longer available so that V03 was the natural choice to provide an extension of 
this work to previous epochs.

Our aim here is not to change the values of the positions given by V03. Instead, we simply applied two
of the filters described in Sect.~4 to somewhat homogenise their 
positions\footnote{Available at http://www.imcce.fr/hosted\_sites/saimirror/burpomaf.htm.}
and ours. Procedures equivalent to the other filters were taken into consideration by these authors 
when preparing their data.
We also provide a brief analysis of the respective filtered positions.
 
In this context, we applied the filter based on the distance from the centre of Uranus (Perkin-Elmer
telescope) and then the filter based on the relative 
positions of Oberon with respect to those of the other satellites. 

For a first verification, however, the data reported in V03 were separated into two sets: photographic (1988 and older) and CCD 
(after 1988) observations. The results are presented 
in the same way as for Table~\ref{table9} in Table~\ref{table12}. The filter related to the distance 
from the centre of Uranus only eliminated Miranda data (62 positions) because the satellites were 
observed when their orbits were mostly face-on.

\begin{table}
\caption{Overall residuals with respect to Oberon from V03}             
\label{table12}      
\begin{center}          
\begin{tabular}{l c c c c c c}     
\hline\hline
\multicolumn{7}{c}{Photographic}\\
\hline
Satellite & $[\Delta\alpha*]$ & $[\Delta\delta]$ & $\sigma_{\alpha}*$ & $\sigma_{\delta}$ 
&\multicolumn{2}{c}{\# positions}\\
  & 
\multicolumn{2}{c}{(mas)}& 
\multicolumn{2}{c}{(mas)}&
before & after \\
\hline     
Miranda  & -32 & -15 & 164 & 118 & 315 & 305\\
Ariel & -7 & -12 & 66 & 55 & 363 & 340\\
Umbriel & -11 & -4 & 73 & 60 & 363 & 335\\
Titania & -1 & -8 & 52 & 44 & 363 & 337\\
\hline
\hline
\multicolumn{7}{c}{CCD}\\
\hline     
Miranda  & 1 & 6 & 71 & 72 & 1130 & 897\\
Ariel & 4 & 12 & 39 & 42 & 1360 & 1104\\
Umbriel & 3 & 10 & 37 & 51 & 1350 & 1113\\
Titania & -10 & 17 & 52 & 61 & 1365 & 1214\\
\hline\hline       
\end{tabular}
\end{center}
Filters described
in steps 1 and 4, in this order (see Sect.~4), were applied separately to photographic and CCD data.
Columns 1 to 5 have the same meaning as those in Table~\ref{table9}. Columns
6 and 7 show the number of measurements before and after applying the
second filter (based on the relative positions of Oberon with respect to those of the other satellites).
Ephemeris positions were obtained from DE432$+$ura111.
\end{table}

It is expected, in Table~\ref{table12}, that the standard deviations for the photographic measurements
be larger than those made with CCD. In addition, since this work profited from a better
reference catalogue for astrometry, we consider that the efforts
of V03 to determine relative angular 
measurements for the satellites were efficient, although the standard deviations in Table~\ref{table12}
are larger than those in Table~\ref{table9}.

In this context, we consider that both photographic and CCD positions are a homogeneous set of data, 
despite the much larger uncertainties of  the Miranda photographic positions. From this point on, 
we therefore emphasise that the filter based on the relative positions of Oberon with respect to those of the 
other satellites is applied without 
distinguishing between the detectors for the V03 data. This slightly changes the total number 
(CCD plus photographic) of positions for each satellite as given by the last column of 
Table~\ref{table12}.

\begin{table}
\caption{Comparison  between the positions reported by V03 and those from this work}             
\label{table13}      
\begin{center}          
\begin{tabular}{l c c c c c}     
\hline\hline
\multicolumn{6}{c}{Original offsets in V03}\\
\hline
Satellite & $\Delta\alpha*$ & $\Delta\delta$ & $\sigma_{\alpha}*$ & $\sigma_{\delta}$ 
&\# positions\\
& 
\multicolumn{2}{c}{(mas)}& 
\multicolumn{2}{c}{(mas)}& \\
\hline     
Miranda  & -119 & -28 & 114 & 131 & 326\\
Ariel & -128 & -38 & 102 & 123 & 469\\
Umbriel & -120 & -46 & 100 & 117 & 474\\
Titania & -136 & -44 & 107 & 115 & 513\\
Oberon & -127 & -58 & 106 & 133 & 551\\
\hline
\hline
\multicolumn{6}{c}{V03 minus this work}\\
\hline     
Miranda  & -22 & 17 & 102 & 133 & 326\\
Ariel & -34 & 21 & 87 & 120 & 469\\
Umbriel & -33 & 12 & 90 & 115 & 474\\
Titania & -45 & 17 & 97 & 116 & 513\\
Oberon & -32 & 0 & 89 & 130 & 551\\
\hline\hline       
\end{tabular}
\end{center}
Offsets from V03 (top table) and offsets from the difference in the sense positions from V03 minus
(common) positions from this work (bottom table).
\end{table}

Table~\ref{table13} compares the original, filtered  positions from V03, with those obtained in 
this work. We note that V03 is closer to our results than to those of the ephemerides they
used for comparison (DE403 and GUST86). 

It is worth mentioning that V03 used the USNO-A2.0 \citepads{1998usno.book.....M} catalogue as reference for 
astrometry and applied corrections on it to minimise systematic effects on its positions.
It is known from \citetads{2001ApJ...552..380A} that these effects are larger (in absolute value) than the
offsets between V03 and this work (see Table~\ref{table13}). Therefore, the correction 
that V03 applied on the USNO-A2.0, as expected, brought the celestial frame originally represented by that 
catalogue closer to that represented by the UCAC4. Most probably, smaller offsets and standard deviations
would have been seen  in Table~\ref{table13} (V03 minus this work section) if proper motions were available to all reference stars used by
V03. These authors reported geocentric positions and we changed our positions
accordingly for these comparisons.

\begin{figure}
   \centering{
   \includegraphics[width=6.5cm]{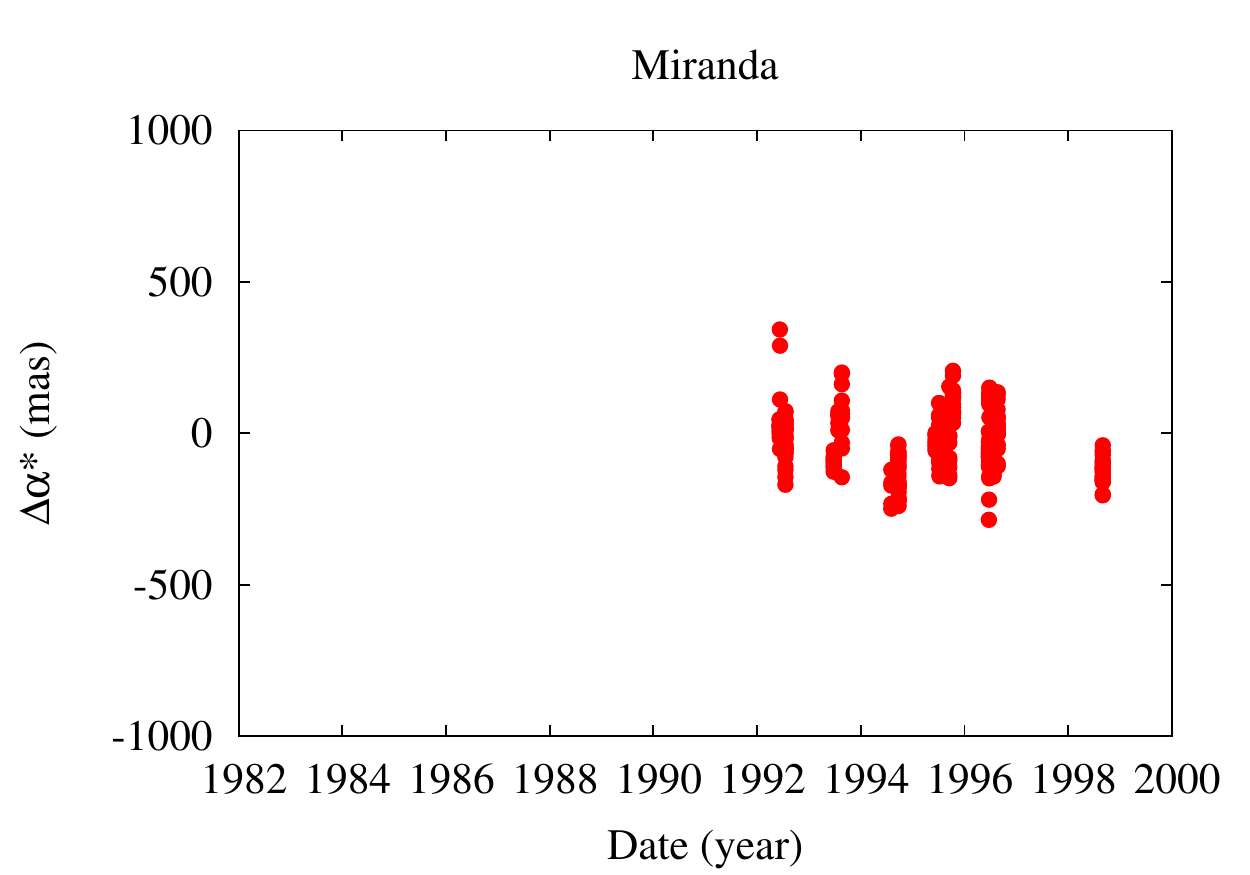}
   \includegraphics[width=6.5cm]{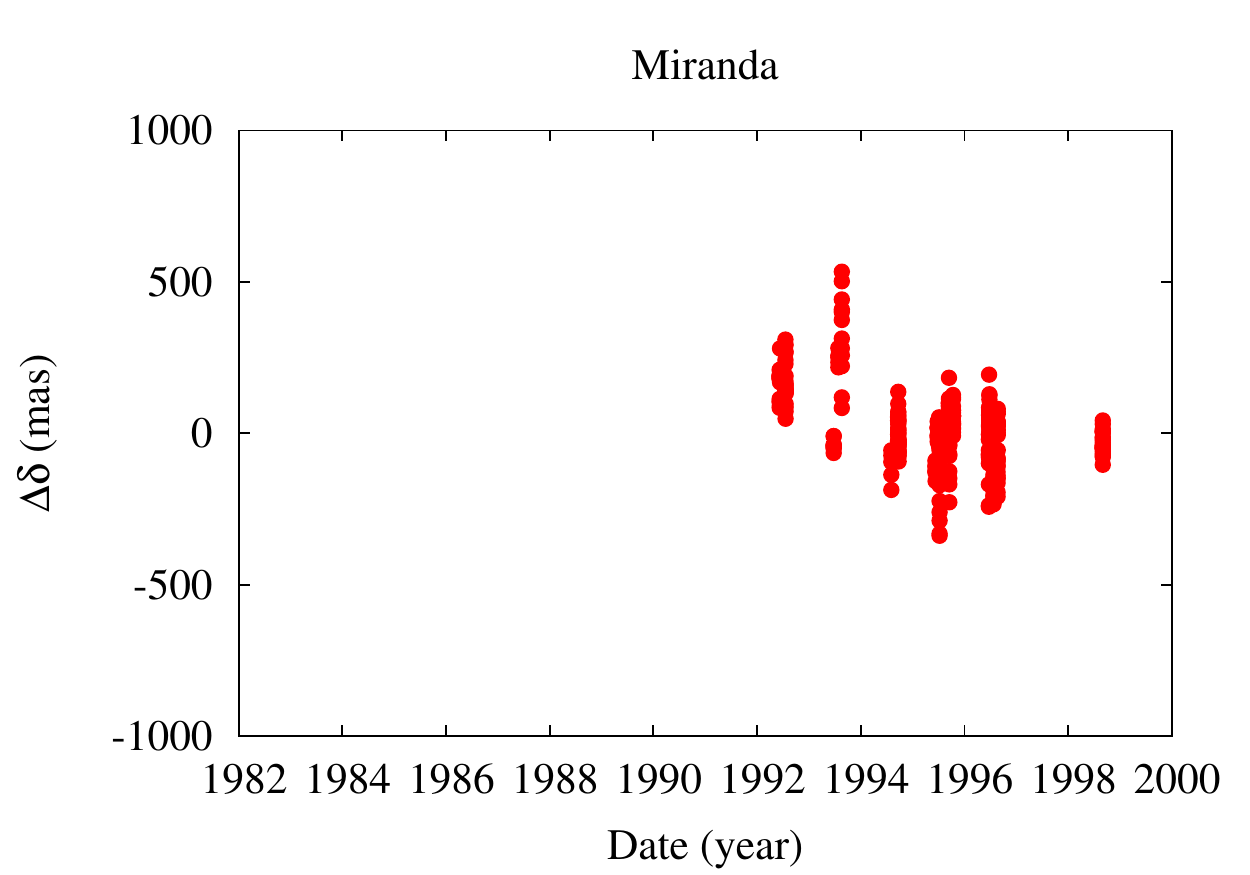}}     
\caption{Updated offsets of V03 for Miranda as a function of time.
             }
         \label{figure17}
   \end{figure}

\begin{figure}
   \centering{
   \includegraphics[width=6.5cm]{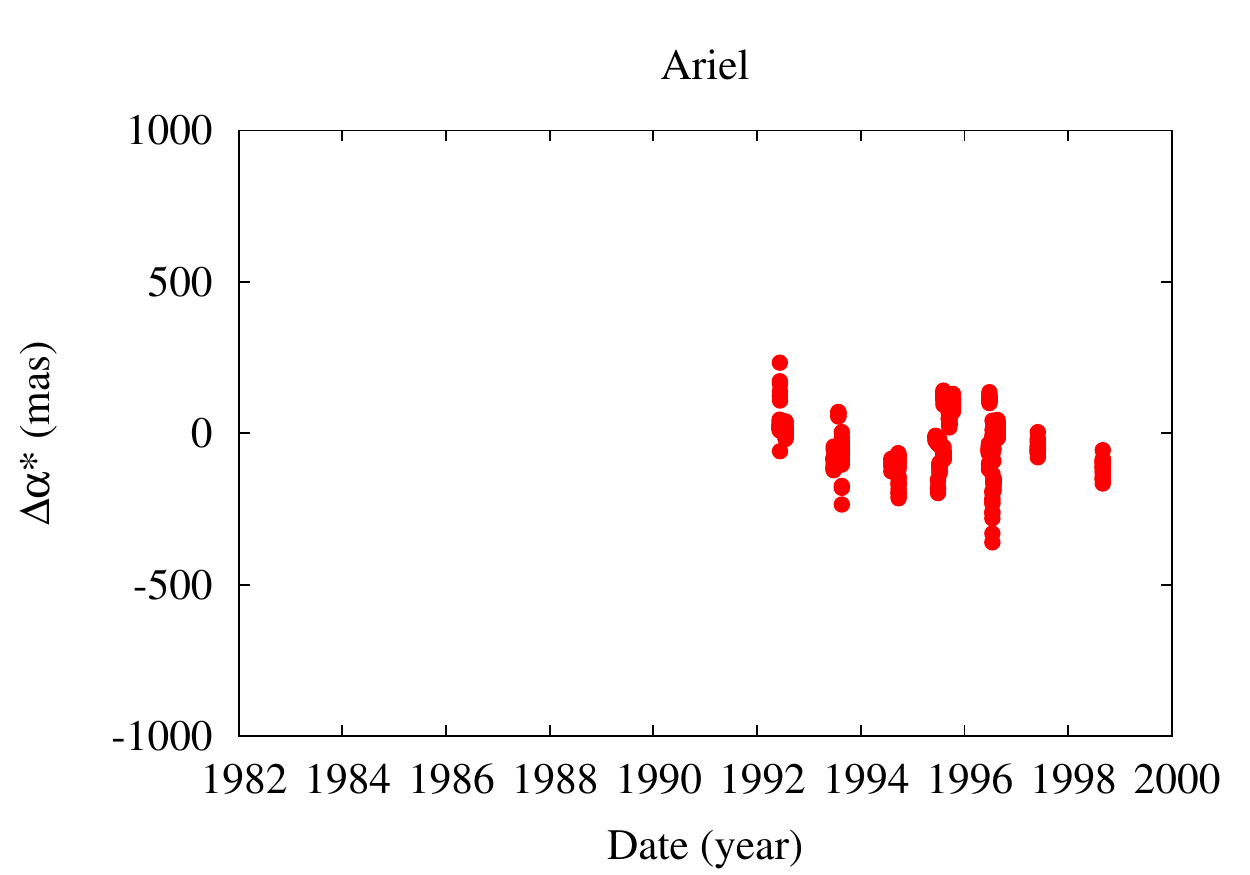}
   \includegraphics[width=6.5cm]{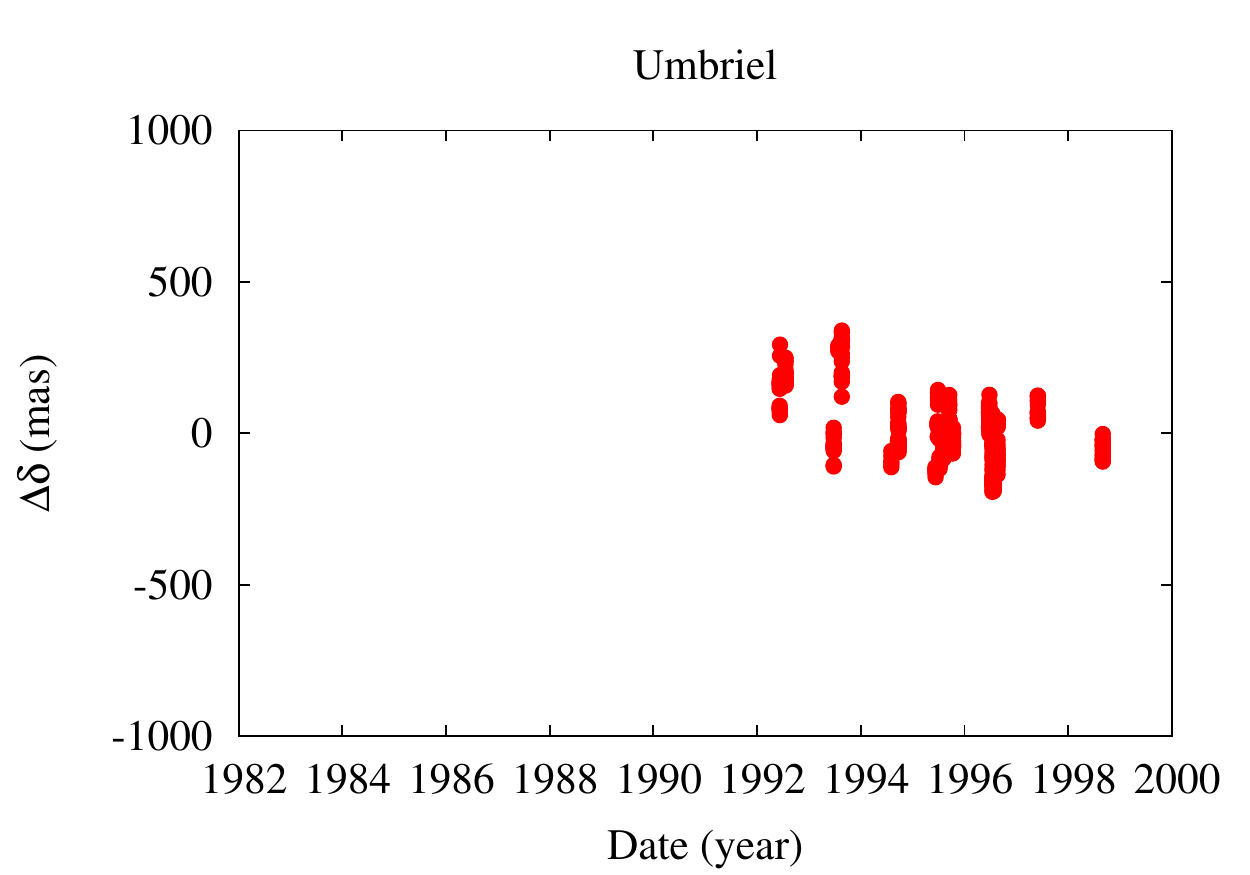}}   
\caption{Updated offsets of V03 for Ariel as a function of time.
             }
         \label{figure18}
   \end{figure}

\begin{figure}
   \centering{
   \includegraphics[width=6.5cm]{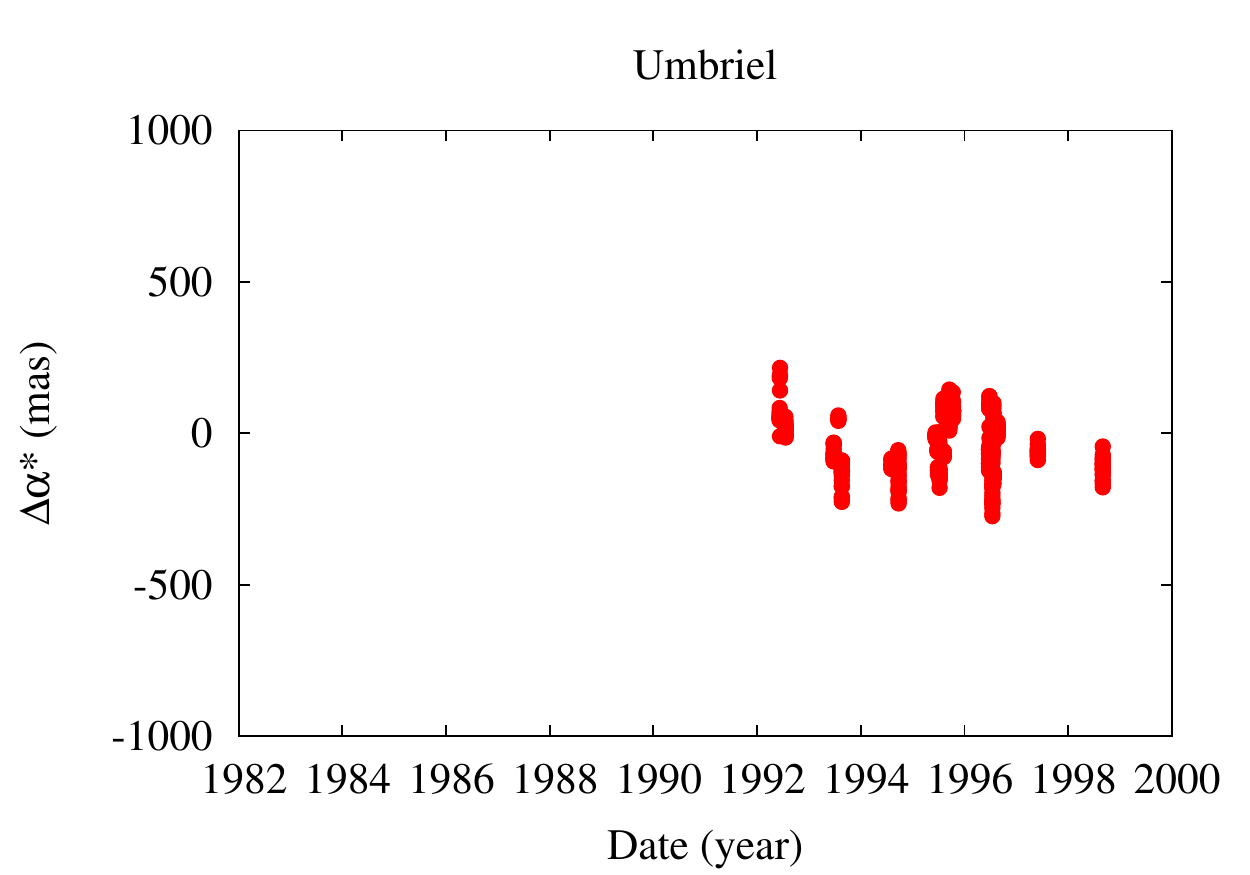}
   \includegraphics[width=6.5cm]{26385_ap_figure19b}}      
\caption{Updated offsets of V03 for Umbriel as a function of time.
             }
         \label{figure19}
   \end{figure}

\begin{figure}
   \centering{
   \includegraphics[width=6.5cm]{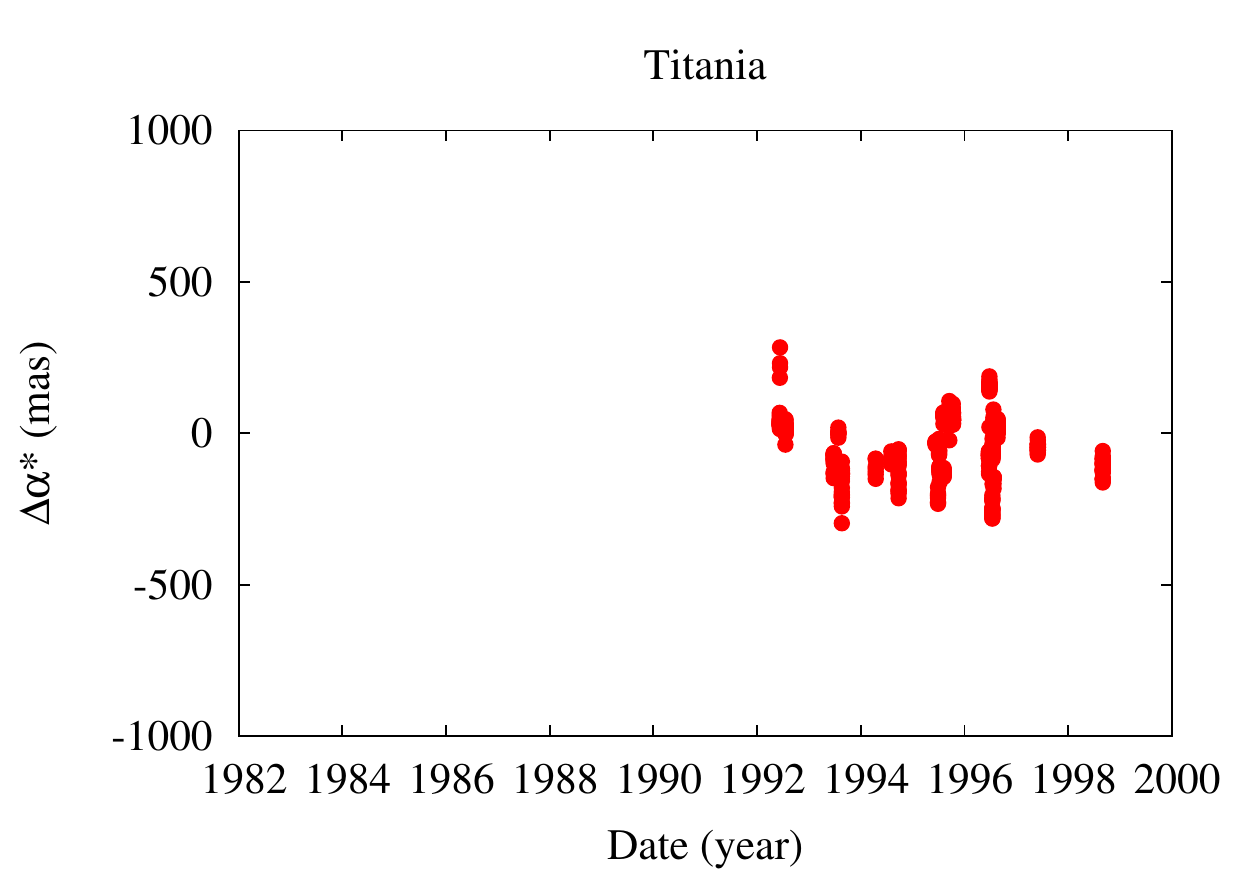}
   \includegraphics[width=6.5cm]{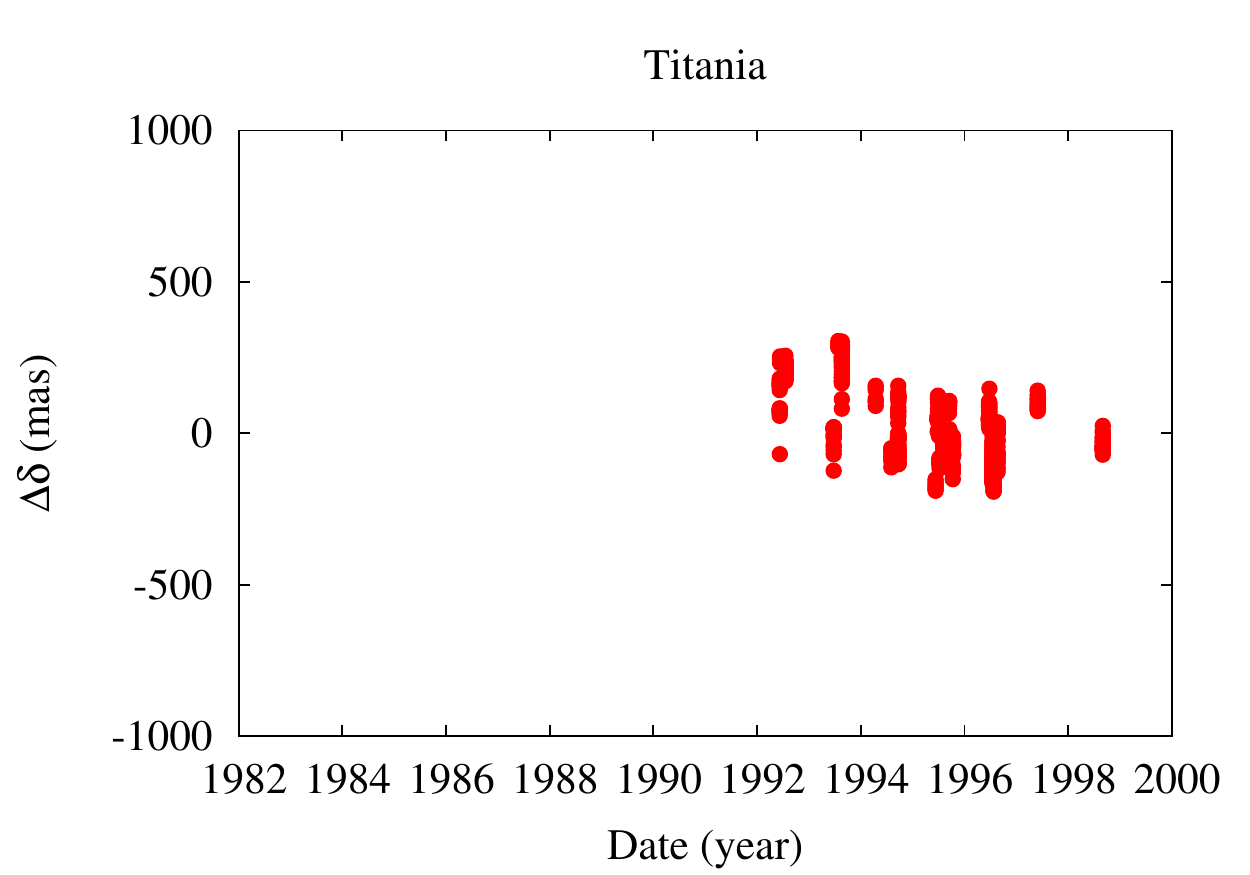}}      
\caption{Updated offsets of V03 for Titania as a function of time.
             }
         \label{figure20}
   \end{figure}

\begin{figure}
   \centering{
   \includegraphics[width=6.5cm]{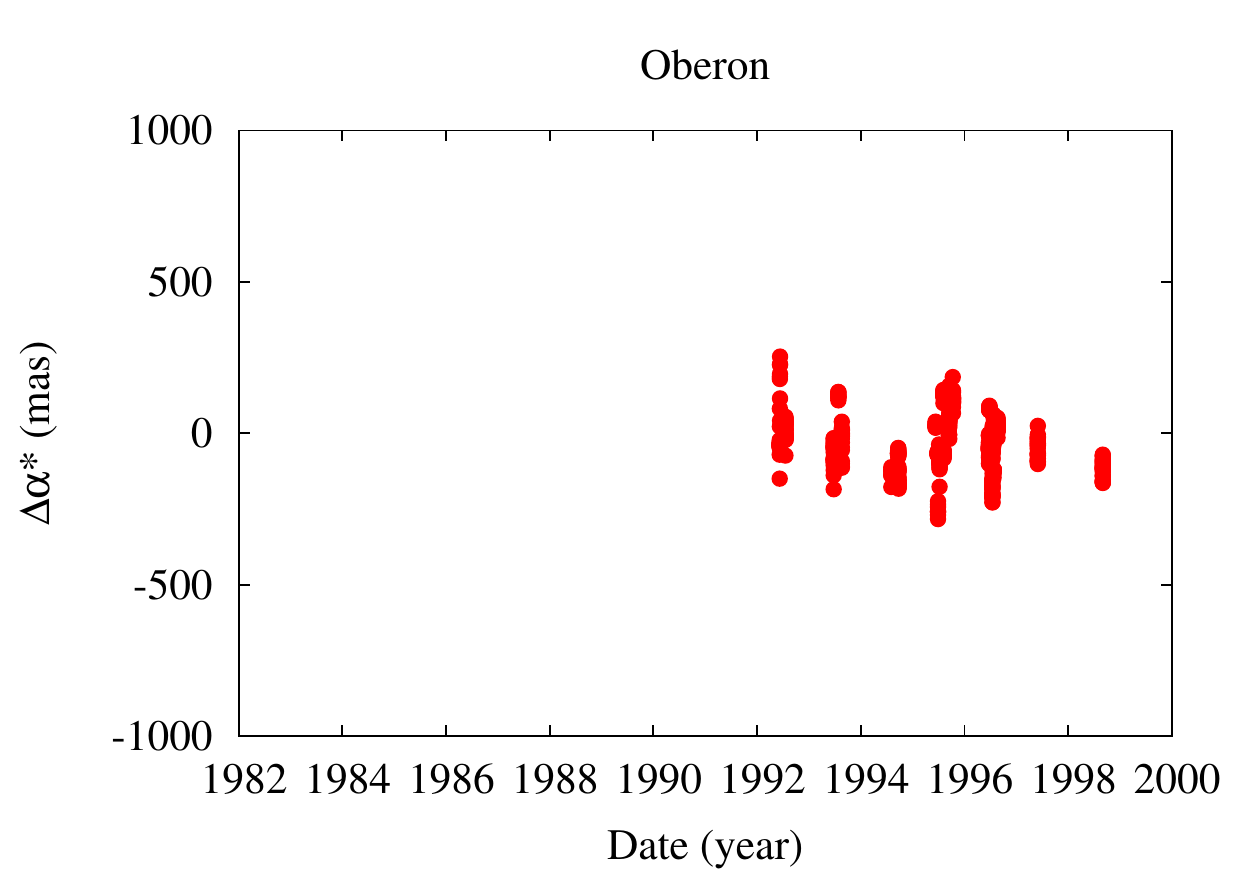}
   \includegraphics[width=6.5cm]{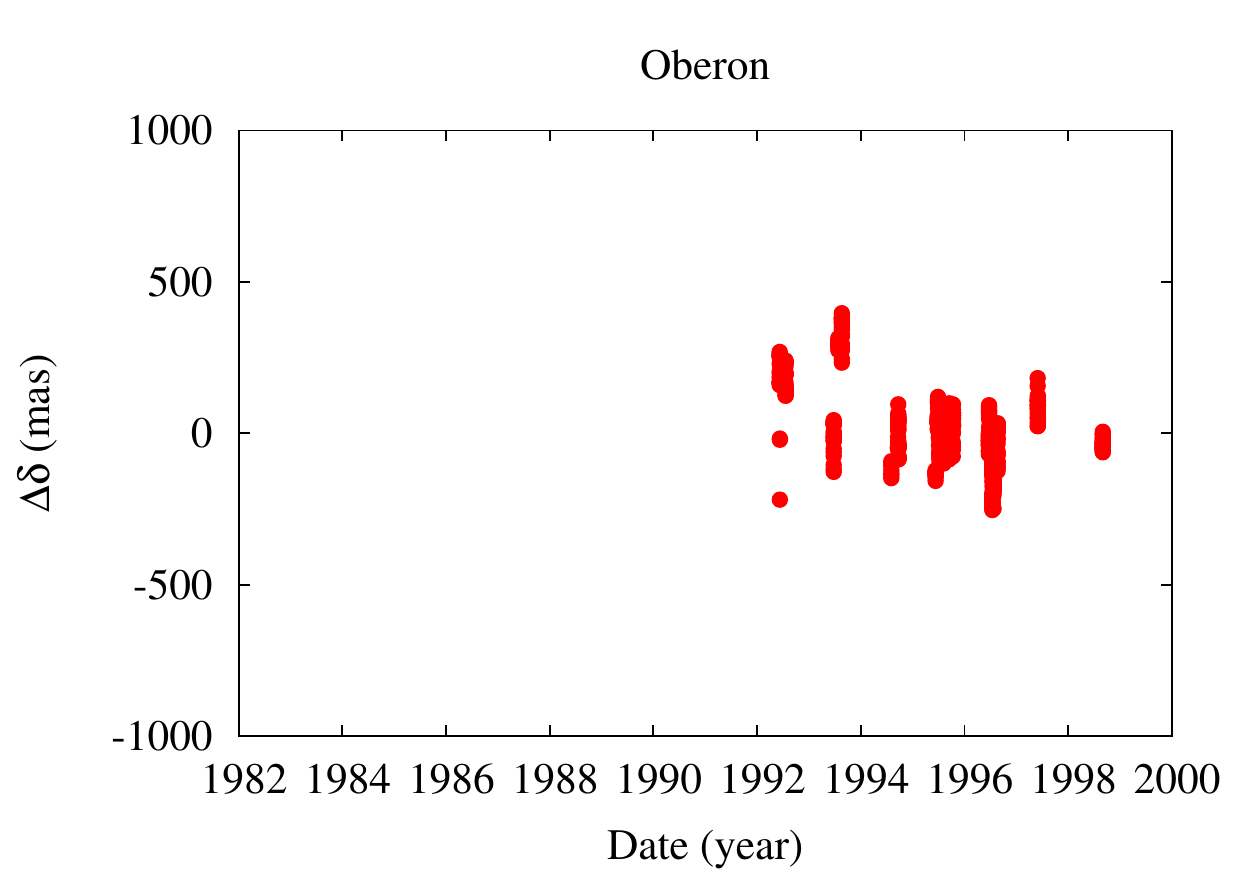}}     
\caption{Updated offsets of V03 for Oberon as a function of time.
             }
         \label{figure21}
   \end{figure}

Figures~\ref{figure17} to \ref{figure21} show the differences in the sense V03 minus this work as a 
function of time. The dimensions of the axes in these figures facilitate the comparison 
with the similar axes in V03.

\section{Position catalogues: Uranus and its main satellites}

Catalogues containing all the observed positions of the five main satellites and Uranus, as
well as the X and Y CCD coordinates of the observed satellites and reference stars, are only 
available at the Centre de Donn\'ees astronomiques de Strasbourg (CDS). 
Table~\ref{table14} shows an extract of one such catalogue of positions for Oberon. Positions of Uranus (see
Table~\ref{table15} for an extract) are also provided. Total counts are provided in
Table~\ref{table16}.

\begin{table}
\caption{Positions of Oberon - extract}             
{\small
\label{table14}      
\begin{center}          
\begin{tabular}{c c c r r}     
\hline\hline       
 JD & \multicolumn{2}{c}{R.A.\hskip15 pt (ICRS)\hskip 15pt DEC.} & 
$\sigma_{\alpha}*$ & $\sigma_{\delta}$ \\
(UTC) & \multicolumn{2}{l}{(h\hskip 7pt m\hskip 5pt s)\hskip 37pt 
($^{\rm o}$\hskip 10pt $^{\prime}$\hskip 8pt $^{\prime\prime}$)} &
\multicolumn{2}{c}{(mas)} \\
\hline     
2448782.68756100 & 19 14 54.182 & -22 45 53.11 & 57 & 49 \\
2448782.70138981 & 19 14 54.042 & -22 45 53.33 & 56 & 48 \\
2448782.73052454 & 19 14 53.747 & -22 45 53.75 & 65 & 56 \\
$\vdots$ & & & & \\
\hline\hline       
\end{tabular}
\end{center}
}
Columns are Julian date (UTC) of the observation; observed ICRS right ascension; 
observed ICRS declination; uncertainty in right ascension - note the term cos$\delta$; uncertainty in 
declination. For the positions of Oberon and the other satellites from V03, two
extra columns provide the values to transfer their original geocentric positions
to the topocenter. See text for details.
\end{table}

Since Uranus is saturated in most of the images, its 
positions (see Table~\ref{table15}) were not measured. They were derived by applying the mean offset on the ephemeris
position of Uranus as obtained from the 
satellites in the same image (except for Miranda). We stress that the positions of Uranus
provided in this paper are not observed ones.

We also note that ephemeris positions of Uranus and the 
offsets for the satellites are both determined from DE432$+$ura111 (this is
also valid when we determine positions of Uranus from V03 data). Taking into 
consideration that the positions in Table~\ref{table15} were most frequently derived 
from the offsets of two to four satellites, discrepancies
in ura111 (for instance, Figs.~\ref{figure12} to \ref{figure16}) are attenuated and,
therefore, the derived positions of Uranus can be regarded as ephemeris independent.

Positions from V03  from before 1992 for Uranus and the main satellites are provided in 
separate catalogues (and separately from our data) with the same format  as shown in 
Tables~\ref{table14} and \ref{table15}, respectively. They are filtered by the 
procedure described in Sect.~6 and presented in their original form, that is, the right ascensions and declinations
are exactly those provided by their paper.
The application of the 
offsets given by the lower part of Table~\ref{table13} on V03 positions is a way to improve their results.
We also note that two extra columns are provided in the tables, containing the data from V03. They
give the values $\Delta\alpha{\rm cos}\delta$ and $\Delta\delta$, in units of milliarcseconds, that should
be added to the right ascension and declination, respectively, to transfer these positions from the
geocenter to the topocenter.

The observational data in V03 span from 1982 to 1998, so that all positions we provide after 1998 (5,827 for the five satellites) are certainly unprecedented. 

We stress that our positions are topocentric,
whereas the original positions of V03 are geocentric.
 
\begin{table}
\caption{Positions of Uranus - extract}   
{\tiny        
\label{table15}      
\begin{center}          
\begin{tabular}{c c c r r r}     
\hline\hline       
 JD & \multicolumn{2}{c}{R.A.\hskip15 pt (ICRS)\hskip 15pt DEC.} & 
$\sigma_{\alpha}*$ & $\sigma_{\delta}$ & sat. \\
(UTC) & \multicolumn{2}{l}{(h\hskip 7pt m\hskip 5pt s)\hskip 37pt 
($^{\rm o}$\hskip 10pt $^{\prime}$\hskip 8pt $^{\prime\prime}$)} &
\multicolumn{2}{c}{(mas)}& \\
\hline     
2448782.68756100 & 19 14 54.553 & -22 45 10.28 & 57 & 49 & o    \\
2448782.70138981 & 19 14 54.431 & -22 45 10.51 & 47 & 42 & tauo \\
2448782.73052454 & 19 14 54.171 & -22 45 10.99 & 58 & 51 & to   \\
$\vdots$ & & & & & \\
\hline\hline       
\end{tabular}
\end{center}
}
Columns are Julian date (UTC) of the observation; observed ICRS right ascension; 
observed ICRS declination; uncertainty in right ascension - note the term cos$\delta$; uncertainty in 
declination; satellites that contributed to the determination of Uranus' position: (a) - Ariel, (u) - Umbriel,
(t) - Titania, (o) - Oberon. For the positions of Uranus from V03, two
extra columns provide the values to transfer their original geocentric positions
to the topocenter. See text for details.
\end{table}

\begin{table}
\caption{Total counts for the catalogues}             
\label{table16}      
\begin{center}          
\begin{tabular}{l c c}     
\hline\hline
Object & \multicolumn{2}{c}{\# positions}\\
  & 
This work (from 1992) & V03 (before 1992)\\
\hline     
Miranda  & 584 & 476\\
Ariel & 1710 & 474\\
Umbriel & 1987 & 462\\
Titania & 2588 & 558\\
Oberon & 2928 & 732\\
Uranus & 3516 & 732 \\
\hline\hline       
\end{tabular}
\end{center}
Number of positions for each satellite and Uranus.
\end{table}

\subsection{Uncertainties in position}

Uncertainties in position (see Tables~\ref{table14} and \ref{table15}) are given by
\begin{equation}
\sigma_{\alpha*,\delta}=\sqrt{\sigma_{a}^{2}+\sigma_{b}^{2}}.
\end{equation}

For all satellites whose positions were determined in this work, 
$\sigma_{a}$ is the standard deviation in right ascension or declination of the reference stars in the 
image containing the respective
satellite and $\sigma_{b}$ is read from Cols. 4 (right ascension) or 5 (declination) in
Table~\ref{table9} for the corresponding satellite. For Oberon, the values of 32 mas (right
ascension) or 29 mas (declination) were used for $\sigma_{b}$. They represent the mean values of the 
uncertainties in Table~\ref{table9} for Ariel, Umbriel, and Titania.
For satellite positions
from V03, uncertainties were determined by taking $\sigma_{a}$=100 mas (right ascension
and declination), and $\sigma_{b}$ is read from Cols. 4 (right ascension) or 5 (declination)
in Table~\ref{table12} for the corresponding satellite. For
Oberon, the values of 43 mas (right ascension) and 51 mas (declination) were used for $\sigma_{b}$.
They represent the mean values of the uncertainties in Table~\ref{table12} (CCD section, where most
of the positions are found) for Ariel, 
Umbriel, and Titania.
The value of 100 mas for $\sigma_{a}$ is a tentative agreement between Cols. 4 and 5 in 
Table~\ref{table13} (V03 minus this work section), in an attempt to reproduce the standard deviation 
of the reference stars in V03.

For all positions of Uranus determined in this work, 
$\sigma_{a}$ is the standard deviation of the reference stars in right ascension or declination
in the respective image containing Uranus, and $\sigma_{b}$ is the standard deviation of the satellite offsets
in either right ascension or declination, used to correct the ephemeris position of Uranus. When only one satellite was used, 32 mas (right ascension) or 29 mas (declination) were used for 
$\sigma_{b}$, see the reasoning given above for these values.
For positions of Uranus from V03, uncertainties were determined by taking again $\sigma_{a}$=100 mas 
(right ascension and declination) and $\sigma_{b}$ as the standard deviation of the offsets of the satellites,
in either right ascension or declination, used to correct the ephemeris position of Uranus. When only one satellite was used, the values of 43 mas (right ascension) or 51 mas (declination) were
taken, see the reasoning given above for these values.


\section{Conclusions and comments}

      We determined accurate positions of the five main satellites of Uranus: Miranda, 
Ariel, Umbriel, Titania, and Oberon for the time span 1992 -- 2011 from observations made at the Pico dos
Dias Observatory. Positions of Uranus, derived from those of the satellites, were also determined. The 
standard deviation of the offsets is typically between 40 mas and 65 mas.
      
      Our positions contribute to the general knowledge of the physics of the Uranus system \citepads[see][]{2014AJ....148...76J} to improve modern planetary ephemerides, such as DE432 and
INPOP13c, as well as modern models of the satellite motions, such as ura111 and NOE-7-2013.
     
      A number of observations made far from the opposition are certainly useful to improve the
determination of the heliocentric distance of Uranus.
     
      The positions from V03 can be used to provide an extension to previous epochs of our
data. The application of the offsets provided by the lower section of Table~\ref{table13} on their original 
positions is a way to improve them.

\begin{acknowledgements}
J.I.B. Camargo acknowledges CNPq for a PQ2 fellowship (process number 308489/2013-6).
F.P. Magalh\~aes acknowledges a CAPES fellowship (process number 657943) for a MSc. at 
Observat\'orio Nacional. RVM acknowledges grants CNPq-306885/2013, Capes/Cofecub-2506/2015, and  Faperj/PAPDRJ-45/2013.
M.A. acknowledges CNPq grants 473002/2013-2, 482080/2009-4 and 308721/2011-0, and FAPERJ grant 111.488/2013.
F.B.R. acknowledges PAPDRJ-FAPERJ/CAPES E-43/2013 number 144997, E-26/101.375/2014.
ARGJ thanks for the financial support of CAPES.
Software routines from the IAU SOFA Collection were used. Copyright \copyright~
International Astronomical Union Standards of Fundamental Astronomy (http://www.iausofa.org).
NOE-7-2013-MAIN ephemeris and data are issued from the activity of the Bureau des Longitudes in Celestial 
Mechanics and Astrometry. It was supported by the ESPaCE (European Satellite Partnership for Computing Ephemerides) 
FP7-Program under ESA grant agreement contract 263466. 
All softwares we used from SOFA, SPICE, and NOVAS are
in FORTRAN.
\end{acknowledgements}


\bibliographystyle{aa}
\bibliography{26385_ap}
\end{document}